\documentclass[12pt,preprint]{aastex}

\usepackage{natbib}
\newcommand{\lsim}{\
\raise-2.truept\hbox{\rlap{\hbox{$\sim$}}\raise5.truept\hbox{$<$}\ }}
\newcommand{\gsim}{\
\raise-2.truept\hbox{\rlap{\hbox{$\sim$}}\raise5.truept\hbox{$>$}\ }}

\shorttitle{SBF for distances and stellar populations studies}
\shortauthors{Cantiello et al.}

\begin{document}

\title{Surface Brightness Fluctuations from archival ACS images: 
a stellar population and distance study\altaffilmark{1}}

\author{Cantiello, Michele\altaffilmark{2,3}}
\author{Blakeslee, John P.\altaffilmark{2}}
\author{Raimondo, Gabriella\altaffilmark{3}}
\author{Brocato, Enzo\altaffilmark{3}}
\author{Capaccioli, Massimo\altaffilmark{4,5}}

\altaffiltext{1}{Based on observations made with the NASA/ESA Hubble
Space Telescope, obtained from the Data Archive at the Space
Telescope Science Institute, which is operated by the Association of
Universities for Research in Astronomy, Inc., under NASA contract NAS
5-26555. These observations are associated with program \#10642.}
\altaffiltext{2}{Department of Physics and Astronomy, Washington State University,
Pullman, WA 99164;  email: jblakes@wsu.edu}
\altaffiltext{3}{INAF--Osservatorio Astronomico di
Teramo, Via M. Maggini, I-64100 Teramo, Italy; email: brocato, cantiello, raimondo@oa-teramo.inaf.it}
\altaffiltext{4}{Dipartimento di Scienze Fisiche, Universit\`a Federico II di
Napoli, Complesso Monte S. Angelo, via Cintia, 80126, Napoli, Italy}
\altaffiltext{5}{INAF--Osservatorio Astronomico di Capodimonte, via
Moiariello 16, 80131 Napoli, Italy; email: mc@na.infn.it }

\begin{abstract}
We derive Surface Brightness Fluctuations (SBF) and integrated
magnitudes in the V- and I-bands using Advanced Camera for Surveys
(ACS) archival data. The sample includes 14 galaxies covering a wide
range of physical properties: morphology, total absolute magnitude,
integrated color. We take advantage of the latter characteristic of
the sample to check existing empirical calibrations of absolute SBF
magnitudes both in the I- and V-passbands. Additionally, by
comparing our SBF and color data with the Teramo-SPoT simple stellar
population models, and other recent sets of population synthesis
models, we discuss the feasibility of stellar population studies
based on fluctuation magnitudes analysis. The main result of this
study is that multiband optical SBF data and integrated colors can be
used to significantly constrain the chemical composition of the
dominant stellar system in the galaxy, but not the age in the case of
systems older than 3 Gyr.

SBF color gradients are also detected and analyzed. These SBF gradient
data, together with other available data, point to the existence of
mass dependent metallicity gradients in galaxies, with the more
massive objects showing a non--negligible SBF versus color
gradient. The comparison with models suggests that such gradients
imply more metal rich stellar populations in the galaxies' inner
regions with respect to the outer ones.
\end{abstract}


\section{Introduction}

The direct investigation of the evolutionary properties of the stars
in a galaxy relies on the availability of individual stellar
spectro--photometric data. However, detailed resolved star information
is achievable only for a few nearby galaxies, thus the present
understanding of stellar populations properties in external galaxies
is basically founded on unresolved star studies, i.e., integrated
starlight information \citep[e.g.][]{trager06}. Therefore, much effort
has been made to establish new, more powerful instruments, and
observational techniques to restore the information lost in the light
integration.

In the last few years the Surface Brightness Fluctuations (SBF) method
has proved to be a powerful technique for both determining the
distance and to probe the stellar populations in extragalactic
systems.

The theoretical basis of the SBF technique is described in
\citet{ts88}, and \citet{tal90}. The fluctuations in the surface
brightness are due to the Poissonian distribution of unresolved stars
in a galaxy. By definition the SBF is the variance of these
fluctuations, normalized to the local mean flux of the galaxy (after
subtracting a smooth galaxy model). As a consequence of its definition,
the SBF amplitude corresponds to the ratio of the second to the first
moment of the stellar luminosity function. Hence, coupling SBF
magnitudes and colors, with classical integrated magnitudes and colors
gives at the same time informations on the first two moments of the
stellar luminosity function in a galaxy.  Specifically, in consequence
of their definitions, the first moment of the stellar luminosity
function (surface brightness, color) carries information on the most
populated stellar phases, i.e. the Main Sequence, while the SBF is
weighted towards the brightest component of the system, namely Red
Giant Branch, and Asymptotic Giant Branch (RGB, and AGB respectively)
stars.

Taking advantage of these properties, unresolved stellar populations
studies have been presented by several authors using SBF to integrated
magnitudes comparisons. Such analysis is based on the comparison of
data with populations synthesis models. Besides the first attempts to
model the SBF signal using numerical techniques \citep{tal90,
buzzoni93,worthey93a}, more models have been provided by several
groups \citep[][to quote only the most recent
ones]{mouhcine05,raimondo05,marin06} covering a wide range of ages,
chemical compositions, and photometric systems.

The first targets for SBF studies were normal elliptical galaxies, as
they were thought to represent a fairly uniform morphological class.
Nevertheless, the SBF method has been subsequently extended to a
wealth of other sources: bulges of spirals \citep{tonry97}, dwarf
galaxies \citep[][and references therein]{rekola05}, giant
ellipticals, galactic and Magellanic Clouds stellar clusters
\citep{at94,gonzalez04,raimondo05}.

The increase of the family of objects with SBF measurements,
accompanied by the fact that ellipticals are not a homogeneous class
in terms of stellar populations properties, had two main effects.
First, concerning distance studies, a few authors engaged in large
campaigns with the purpose of calibrating the absolute SBF magnitude
against physical properties of galaxies, e.g. the $(V-I)_0$ color, so
that reliable distance estimations can be derived for galaxies with
substantial physical differences \citep{tonry01,mieske06,mei07vii}.

Second, since different classes of objects are on average expected to
host different stellar systems, several groups faced the problem
of revealing/analyzing the physical and chemical attributes of
unresolved stellar populations using the SBF technique. Typically, the
comparison of data with models has been adopted to investigate the
properties of unresolved old stellar systems, by using either SBF
absolute magnitudes, SBF colors, or even SBF gradients
\citep[][]{cantiello03,jensen03,cantiello05}; also resolved stellar
systems  have been analyzed using SBF data \citep[][R05
hereafter]{gonzalez04,raimondo05}.

Taking advantage of the encouraging results offered by the SBF
technique for both distance and stellar populations studies, in this
paper we carry out a multi--color SBF analysis using the rich archive
of HST/ACS data. 

The large format, high resolution, sharp Point Spread Function (PSF),
good sampling, and public access to ACS \citep{ford98,sirianni05}
images obtained for other science goals, make this camera ideal for
SBF archival research studies. As demonstrated by \citet{mei05iv} and
\citet[][C05 hereafter]{cantiello05}, the significant geometric
distortion that affects the ACS images does not represent a major
problem in the measurement of SBF magnitudes -- see the quoted papers
for more details.

We present multi--color SBF and integrated magnitudes measurements for
a sample of galaxies imaged with the ACS. The list of objects included
in our sample covers a wide range of galaxy properties. We take
advantage of this to examine the accuracy of existing SBF absolute
magnitude calibrations. In addition, using recent populations
synthesis models, we discuss the properties of the dominant stellar
populations in the galaxies.

The paper is organized as follows: Section~2 describes the selected
sample of galaxies, together with the selection criteria. The
procedures to derive surface photometry, isophotal analysis, sources
maps and photometry, and SBF magnitudes are presented in Section~3.
In Section~4 the results of our measurements are analyzed. We discuss
separately the two aspects of SBF as distance indicator, and the SBF
as a stellar population tracer. A summary of the work is given in
Section~5.

\section{Observational data}

We have undertaken an analysis of Archival HST data as part of program
AR-10642. We obtained from the HST archive the images of galaxies for
which ACS/Wide Field Camera V (either F555W, or F606W), and I (F814W)
passband data were available. Among these objects we finally selected
the galaxies with exposure times, and surface brightness profiles
properties sufficient to allow SBF measurements with sufficient signal
to noise (S/N$\gsim$7, \citet{blake99}).

The data have been downloaded from the HST archive.  The image
processing, including cosmic--ray rejection, alignment, and final
image combination, is performed with the APSIS ACS data reduction
software \citep{blake03}. The drizzling kernel adopted is the Lancosz3
as suggested by \citet{mei05iv} and C05.

Table \ref{tab_dati} provides the final catalog of galaxies, together
with some useful quantities, and the image exposure times. The table
columns are (1) galaxy name; (2-3) right ascension and declinations
from NED\footnote{http://nedwww.ipac.caltech.edu} (J2000); (4) the
flow-corrected recession velocity based on the local velocity field
model by \citet{mould00} (km s$^{-1}$, from NED); (5) morphological T
type from Leda\footnote{http://leda.univ-lyon1.fr}; (6) Mg$_2$ index
from Leda; (7) velocity dispersion $\sigma$, from Leda; (8) H$_{\beta}$
from the compilation of \citet{jensen03}; (9) total apparent B
magnitude from Leda; (10) B-band extinction; (11)
group distance modulus; (12) bibliographic reference for the distance
modulus; (13) total exposure time in the I-band; (14) total exposure
time for the V-band image.

The distance moduli in the table are derived from the weighted average
distances of galaxies lying in the same group. The distance are
estimated using several different distance indicators (no SBF
distances are used). For NGC\,474 no group distance is available so we
adopt the single distance.

Data are corrected for galactic extinction using the \citet{sfd98}
extinction maps. The extinction ratios, as well as the
transformations from the ACS photometric system to the standard BVRI
filters, follow the \citet{sirianni05} prescriptions.

In the forthcoming sections we will always consider the standardized
SBF and integrated magnitudes, instead of the ACS ones. In our
previous works -- C05 and \citet{cantiello07} -- we have discussed the
reliability of the \citet{sirianni05} equations for the (F435W, F606W,
F814W)--to--(B, V, I) transformations. Here we mention that for the
sample of galaxies with available (V-I)$_0$ color, the average
difference between our standardized colors and those from literature
is $\Delta (V-I)_{0,~this~work-literature}=0.00 \pm 0.03$ -- this
quantity refers to the average difference between this work colors and
the values from \citet[][T01 hereafter]{tonry01} for the galaxies
NGC\,1316, NGC\,1344, NGC\,3610 and NGC\,3923 derived in the same
galaxy regions.

\section{Photometric, Isophotal and SBF analysis}
The SBF data analysis is done following the  procedure
described in C05 and \citet{cantiello07}. For the present analysis, we
used the same techniques and the same software developed in our
previous works.  In the following sections we give a brief summary of
the procedure. A detailed description can be found in the quoted
papers and references therein.

\subsection{Galaxy modeling}
We adopted an iterative process to determine (1) the sky value, (2)
the best fit of the galaxy, and (3) the mask of the sources in the
frames.

A provisional sky value is obtained as the median pixel value in the
CCD corner with the lowest number of counts. This value is subtracted
from the original image and all the obvious sources whose presence
could badly affect the process of galaxy modeling (bright stars,
extended galaxies, dusty regions) are masked out. The gap region
between the two ACS detectors and other detector artifacts are masked
too.

Then, we fit the galaxy isophotes using the IRAF/STSDAS task ELLIPSE,
which is based on the method described by \citet{jedrzejewski87}.
Once the preliminary galaxy model is subtracted from the
sky--subtracted frame, a wealth of faint sources appears. In the
following we refer to all these sources -- mainly globular clusters
(GC), and background galaxies -- as ``external sources''.  A mask of
the external sources is derived from the frame of external
sources obtained with SExtractor (OBJECTS frame), the mask is obtained by convolving
the external sources frame with a gaussian kernel having the same
FWHM of the PSF. The new mask is then fed to ELLIPSE to refit the
galaxy's isophotes.

After the geometric profile of the isophotes has been determined, we
measure the surface brightness profile of the galaxy in regions matching the
ellipticity and position angle of the isophotes. Then, to improve the
estimation of the sky, we fit the surface brightness profiles with a
de Vaucouleurs $r^{1/4}$ profile plus the constant sky offset
\footnote{It is worth emphasizing here that, as in C05, we have performed some
experiments to test the robustness of the procedure of sky estimation.
In particular, we have studied how the assumption of a de Vaucouleurs
$r^{1/4}$ profile instead of a more general Sersic $r^{1/n}$ profile
affects our results. As a result we have found that adopting a Sersic
profile does not alter substantially the integrated color and SBF
values as these quantities agree within uncertainty with the ones
derived using the $r^{1/4}$ profile approximation.}.

The new sky value is then adopted and the whole procedure of galaxy
fitting, source masking, and surface brightness profile analysis is
repeated, until convergence. Usually, for the less luminous galaxies
of our sample, the sky value obtained from the outer corner is a good
estimation of the final sky value. In those cases were the ACS field
of view is completely filled with the galaxy, the final sky counts are
on average $\sim$10\% smaller than the first estimation -- for the six
galaxies in common with \citet{sikkema06} our sky values agree with
their values within the uncertainties, we have on average
$(sky_{this~work}-sky_{Sikkema}) / sky_{this~work} \sim 0.03 \pm
0.06$.

\begin{figure}[!ht]
\begin{center}
\includegraphics[angle=0,scale=.60]{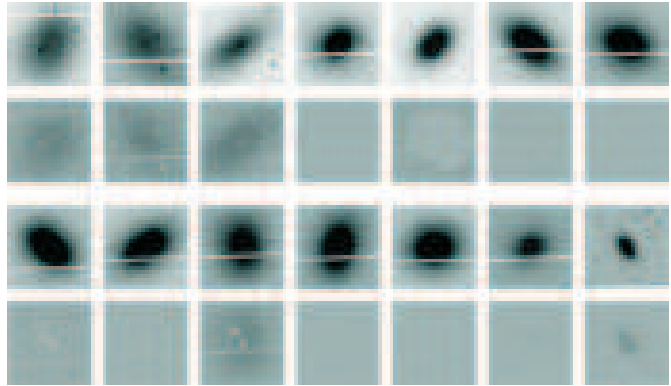}
\caption{The original I-band images and residual frames. The object plotted
are (from left to right, upper rows) DDO\,71, KDG\,61, KDG\,64,
NGC\,474, NGC\,1316, NGC\,1344 and NGC\,2865. Lower rows (left to
right): NGC\,3610, NGC\,3923, NGC\,5237, NGC\,5982, NGC\,7626,
UGC\,7369 and VCC\,941. In the original frames a 10$\arcsec$ segment
is also shown.
\label{fig1}}
\end{center}
\end{figure}

After sky and galaxy model have been subtracted, some large--scale
deviations are present in the frame. We remove these deviations using
the background map (BACK\_SIZE parameter set to 25) obtained running
SExtractor \citep{bertin96} on the sky+galaxy subtracted frame. We
will refer to this sky+galaxy+large--scale residuals subtracted frame
as ``residual frame''. ACS images of the 14 selected galaxies are
shown in Figure \ref{fig1}, together with the residual frames.

We must emphasize that we succeeded in modeling the galaxy light with
elliptical isophotes for all the selected galaxies, including the
objects classified as irregular galaxies (Table \ref{tab_dati}) and
for the galaxies which show prominent shells. This was possible
since $a$) the shells are prominent with respect to the smooth galaxy
profile only in more external regions respect to the ones we
considered for SBF and color measurements (Figure \ref{fig1}); $b$)
the iterative modeling procedure provides a good model of the galaxy
profile; $c$) the large scale residuals subtraction removes most of the
large-scale (shell) features left behind by the modeling.

\subsection{Sources Photometry}

The next step in our procedure is to evaluate the photometric
properties of point--like and extended external sources left in the
image. The construction of the photometric catalog is critical for SBF
measurement, as the estimation of the luminosity function of external
sources is fundamental to properly evaluate the extra fluctuations due
to the undetected sources present in the frame \citep{tal90}.

The photometry of the sources is obtained independently on the I- and
V-band frames using SExtractor, the output catalogs are then matched
using a 0.1$''$ radius. Aperture corrections and extinction
corrections are applied before transforming the ACS magnitudes into
the standard I and V (C05).

The fit of the luminosity function is obtained assuming a gaussian
Luminosity Function for the GC component \citep[GCLF,][]{harris91}:
\begin{equation}
n_{GC}(m) =\frac{N_{0,GC}} {\sqrt{2 \pi \sigma^2}}~~e^{- \frac {(m -
m_{X,GC})^2}{2 \sigma ^2}},
\end{equation}
plus a power-law luminosity function \citep{tyson88} for the background
galaxies:
\begin{equation}
n_{gxy}(m) =N_{0,gxy} 10 ^ {\gamma m},
\end{equation}
where $N_{0,GC}$ \citep[$N_{0,gxy}$,][]{blake95} is the globular
cluster (galaxy) surface density, and $m_{X,GC}$ is the X-band
turnover magnitude of the GCLF at the galaxy distance. In expression
(2) we used the $\gamma$ values obtained by \citet{benitez04}. For the
GCLF we assumed the turnover magnitude and the width of the gaussian
function from \citet{harris01}, i.e -7.4 mag, -8.5 mag for the V-
and I-band turnover magnitudes respectively, and dispersion
$\sigma=1.4$ mag (see also \S 3.3). To fit the total LF we used the
software developed for the SBF distance survey (T01) and optimized for
our purposes; we refer the reader to C05 and references therein for a
detailed description of the procedure. Briefly: a distance modulus
($\mu_{0}$) for the galaxy is adopted in order to derive a first
estimation of $m_{X,GC}=\mu_0+M_{X,GC}$, then an iterative fitting
process is started with the number density of galaxies and GC, and the
galaxy distance allowed to vary until the best values of $N_{0,GC}$,
$N_{0,gxy}$ and $m_{X,GC}$ are found via a maximum likelihood method.

The whole procedure of luminosity function fitting is not applied to
DDO\,71, KDG\,61, KDG\,64, and VCC\,941. For these objects the
few candidate globular clusters (if present) appear resolved in the
ACS images, and are masked out. Similarly, the images of these
galaxies allow us to recognize and mask out most of the brighter
background galaxies (e.g. all sources brighter than 25th
magnitude in the I-band), so that the contribution of the faint
background galaxies is very small compared to the stellar fluctuations
($P_r / (P_0 - P_r) \equiv P_r /P_f$ $<$ 0.001, see Table
\ref{tab_measures} and next paragraph). As a consequence no
extra--fluctuation correction has been applied to these galaxies.

\subsection{SBF measurements}

The pixel--to--pixel variance in the residual image has several
contributors: ($i$) the poissonian fluctuation of the stellar counts
(the signal we are interested in), ($ii$) the galaxy's GC system, ($iii$)
the background galaxies, and ($iv$) the photon and read out noise.

To analyze all such fluctuations left in the residual frame, it is
useful to study the image power spectrum as all the sources of
fluctuation are convolved with the instrumental PSF, except for the
noise. We performed the Fourier analysis of the data with the
IRAF/STSDAS task FOURIER.

In the Fourier domain the photon and read out noise are characterized
by a white power spectrum, thus their contribution to the fluctuations
can be easily recognized as the constant level at high wave numbers in
the image power spectrum.  On the other hand, since the stellar,
globular clusters, and background galaxy fluctuation signals are all
convolved with the PSF in the spatial domain, they multiply the PSF
power spectrum in the Fourier domain. Thus, the total fluctuation
amplitude can be determined as the factor to be multiplied to the PSF
power spectrum to match the power spectrum of the residual frame.

The residual frame, divided by the square root of the galaxy model
\citep{tal90}, is fourier transformed and azimuthally averaged. The
azimuthal average $P(k)$ is used to evaluate the constants $P_0$, and
$P_1$ by fitting the expression:

\noindent
\begin{equation}
P(k)= P_0 \cdot E (k) + P_1 \,,
\end{equation}
\noindent
where $P_1$ is the constant white noise contribution, $P_0$ is the PSF
multiplicative factor that we are looking for, and $E(k)$ is the
azimuthal average of the PSF power spectrum convolved with the mask
power spectrum \citep{tal90}. Since neither contemporary
observations of isolated stars, nor good PSF candidates were available
in our frames, we used the template PSFs from the ACS IDT, constructed
from bright standard star observations.

As mentioned in the previous section, the fluctuation amplitude $P_0$
estimated so far includes stellar fluctuation and the contribution of
unmasked external sources.  To reduce the effect of this spurious
signal, all the sources above a defined signal to noise level
(typically we adopted a $S/N \sim3.5$) have been masked out before
evaluating the residual image power spectrum. Thus, at each radius
from the galaxy center, a well defined faint cutoff magnitude
($m_{lim}$) fixes the magnitude of the faintest objects masked in that
region.  Such masking operation greatly reduces the contribution to
$P_0$ due to the external sources, but the undetected faint and the
unmasked low S/N objects could still affect $P_0$, thus their
contribution -- the residual variance $P_r$ -- must be properly
estimated and subtracted. The residual variance is computed evaluating
the integral of the second moment of the luminosity function in the
flux interval $[0,~f_{lim}]$:

\noindent
\begin{equation}
\sigma_r^2 = \int_0^{f_{lim}} N_{Obj}(f) f^2 df
\end{equation}
\noindent
where $f_{lim}$ is the flux corresponding to $m_{lim}$, and
$N_{Obj}(f)$ is the luminosity function previously obtained as the sum
of the GCLF and the galaxies power law.  The residual variance $P_r$
is then $\sigma_r^2$ normalized by the galaxy surface brightness.
On average, the $P_r$ correction is small for all the galaxies.
Thanks to the faint completeness limit of these images, it is
typically $\sim$5\% (7\%) of the total fluctuations amplitude $P_0$ in
the I (V) band frames, except for the dwarf galaxies for which no
$P_r$ correction has been applied.

Finally, the SBF magnitude is obtained as follows:

\begin{equation}
\bar{m}_X = -2.5 ~ \log (P_0 - P_r) + m_{zero}^{X,ACS} +
2.5~\log(T) - A_{X}
\end{equation}
where X refers to the I or V passband, $m_{zero}^{X,ACS}$ is the
zeropoint ACS magnitude reported by S05 (VEGAMAG system), T is the
exposure time, and A$_{X}$ is the reddening correction.

Since we are also interested in studying the radial behavior of SBF,
the procedure described above is applied to several distinct
elliptical annuli for each galaxy; the annuli shape reflects the
geometry of the isophotes profile.

Table \ref{tab_measures} reports the final results of our measurements
for each annulus and for all galaxies of the sample. Only annuli with
SBF S/N$\geq$7 are taken into account
\citep{jensen96,mei01,cantiello07}.  The table columns give: (1) the
average annular radius in $arcsec$; (2-3) the annulus color and
uncertainty; (4-5) $P_0$ and $P_r$ for the V-band images ; (6-7) the
V-band SBF magnitude; (8-9) $P_0$ and $P_r$ for the I-band images;
(10-11) the I-band SBF magnitude.  For each galaxy also the weighted
average ($\langle av. \rangle_w$) color and SBF magnitudes are
reported, when possible.

The quoted uncertainties are the statistical errors due to the sky
estimations, and the PSF power spectrum fitting. A default $20\%$
uncertainty is associated to $P_r$ \citep{tal90,cantiello07}, and
included in the final SBF error. To test the robustness of the $P_r$
correction versus the sigma fitting parameter adopted for the GCLF, we
have performed some tests by changing the sigmas in the range 0.8 to
1.6 \citep{jordan06}, as a result we find that the effect on the
final SBF magnitudes is $< 0.1$ mag in all cases. All uncertainties
also include the error propagation of the color equations from
\citet{sirianni05}.

Additional systematic errors are $\sim$0.03 mag in the PSF
normalization, $\sim$0.01 mag error from filter zeropoint, and
$\sim$0.01 mag from the flat--fielding -- see C05 for details.

\begin{figure}[!ht]
\plotone{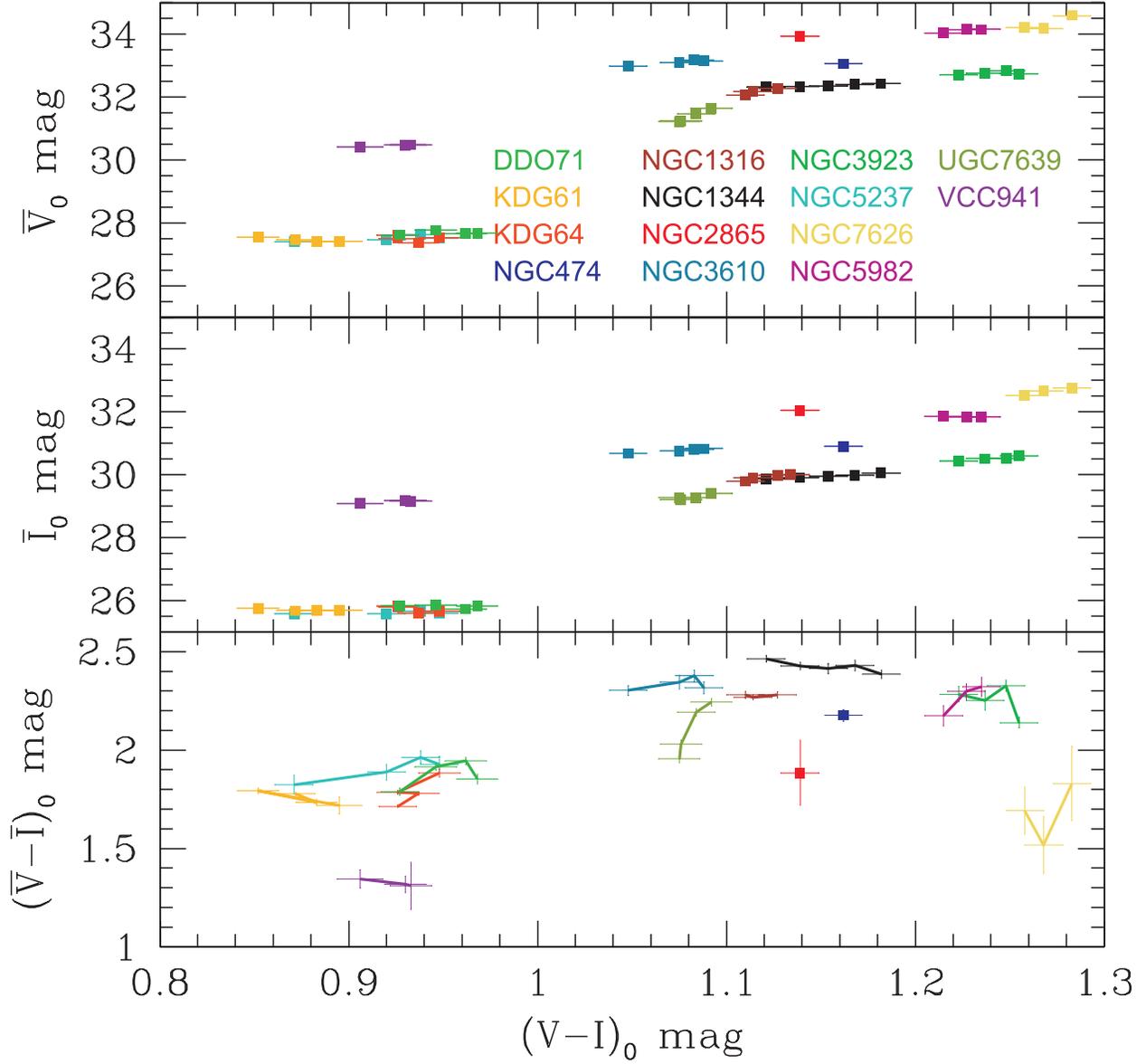}
\caption{SBF apparent magnitudes and color versus integrated (V-I)$_0$
color. Different colors refer to different galaxies as labeled.
\label{sbf_vi}}
\end{figure}

Figure \ref{sbf_vi} shows the V- and I-band SBF apparent magnitudes
and the $(\bar{V}-\bar{I})_0$ annular data versus the annulus
integrated color. As a first impression we see that, unlike our
previous study (C05), no systematic SBF versus color 
gradient feature can be recognized.  The presence of a SBF gradient
has been estimated by fitting a least squares line to the I-band SBF
and (V-I)$_0$ color data\footnote{We have chosen I-band images as
reference due to their higher S/N.}, and comparing the slope,
$\alpha=\delta \bar{m}_I / \delta (V-I)_0$, with its uncertainty.
Since, in all cases the change in color and $\bar{m}_I$ occurs as
function of radius, we indicate in Table \ref{tab_measures} as
``SBF-gradient'' those galaxies with slope $\alpha$ at least twice
bigger than its estimated uncertainty (i.e. $\alpha / \Delta \alpha
\geq 2$), the galaxies which do not fulfill this condition are labeled
as ``SBF-flat''.

\section{Discussion}
As mentioned before, the link between SBF amplitudes and stellar
population properties lies at the base of the determination of the
absolute SBF magnitudes for distances studies, via theoretical or
empirical calibrations, and of the use of SBF to analyze the
properties of the stellar populations in galaxies.

Taking advantage of the characteristics of our set of measurements, in
the following sections we try to answer at two different
questions. First, what can we infer in terms of stellar populations
properties from SBF- and SBF-gradient versus color analysis?  Second, given
the number of calibrations available in literature, which one, applied
to our measurements, gives the best results, i.e. distance moduli in
agreement with the group distances reported in Table \ref{tab_dati}?

\subsection{Stellar population properties}
\subsubsection{Population properties from observational data}

One major difference of the present data from most of the SBF works,
is that the new dataset provides a sample of galaxies covering a wide
range physical properties (total magnitude, morphological type);
as a consequence we expect to find different stellar components to
dominate the light emitted by these objects. Figure \ref{single} shows
a comparison of some galaxy's physical quantities, from Table
\ref{tab_dati}, versus the average integrated color from Table
\ref{tab_measures}. As readily apparent, large differences exist
between the various objects, in particular the sample spans a range of
10 magnitudes in absolute B-band magnitude M$_{Bt}$.

By inspecting the panels of Figure \ref{single}, we note that the
sample can be split in two subsamples: a blue subsample at
(V-I)$_{0}\leq 1.0$ mag, and a red subsample at (V-I)$_{0}> 1.0$ mag. The blue
galaxies, with an average (V-I)$_{0}\sim 0.92$ mag, are generally 
less luminous (M$_{Bt}\sim -13.0 \pm 1.1$) and cover a large range of
morphological types. On the contrary, the red galaxies, at average
(V-I)$_{0}\sim 1.16$ mag, are brighter (M$_{Bt}\sim -20.5 \pm 1.8$),
and morphologically more uniform.

Keeping in mind these points, in this section we take advantage of the
SBF and color data reported in Table \ref{tab_measures} to infer
information on the physical properties of the dominant stellar system
in the galaxies of our sample. We begin our comparisons by considering
the average SBF measurements. Hence, only observational
quantities are considered. We will introduce the comparison with
models, and discuss the caveats of such discussion, later on.

As a first step, we plot the galaxy's physical quantities used in
Figure \ref{single} against the SBF absolute magnitudes
(Fig. \ref{singlesbf}, panels $a$ and $b$) as derived by assuming the
distance modulus quoted in Table \ref{tab_dati}. In the ($b$) panels
the C05 data are also included. As a general result the plots disclose
that SBF magnitudes show trends which are well consistent with what is
found with integrated colors (Fig. \ref{single}). Clearly, the
relationships are far from being well established due to the small
number of galaxies in the sample. In spite of this, the $\bar{V}$ and
$\bar{I}$ $versus$ M$_{Bt}$ panels show a quite defined correlation.

For most of the blue galaxies, no H$_{\beta}$, Mg$_2$, or velocity
dispersion $\sigma$ estimation is available from literature. However,
taking into account the extrapolations shown by the dashed lines
in the panels of Figure \ref{single}, these galaxies are also expected
to have smaller Mg$_2$ and velocity dispersion, and higher H$_{\beta}$
values with respect to the red counterpart. All these general
properties agree with a scenario where the blue objects are low mass
galaxies (low M$_{Bt}$ and $\sigma$), with a relatively young/metal
poor dominant stellar component (low Mg$_2$, high H$_{\beta}$), while
the red objects are expected to be mostly massive galaxies populated
by old, metal rich stellar systems -- see for example
\citet{gallazzi06} for a discussion based on a large sample of
galaxies.

Somewhat more interesting are the correlations shown in the panels
($c$) of Figure \ref{singlesbf}, where we plot the same physical
quantities as function of the SBF color. These measurements have the
relevant feature of being distance-free. Contrary to the absolute SBF
magnitudes, the SBF color shows little or no correlation with the
plotted physical quantities. For example, the lower-left panel shows
that $(\bar{V}-\bar{I})_0$ SBF color does not have a strong
correlation with M$_{Bt}$ absolute magnitude. This is expected on
theoretical basis \citep[e.g.][\S 5.3]{cantiello03}, however this
is the first time that such behavior is explicitly shown.

A further insight of these results is obtained when the presence or
the absence of a radial SBF gradient is taken into account. The
galaxies listed in Table \ref{tab_measures} have been accordingly
divided in two classes: $i)$ ``SBF-gradient'' galaxies are the ones
showing a radial SBF gradient and $ii)$ ``SBF-flat'' galaxies
which show nearly constant SBF magnitudes (within the uncertainties)
over the explored annuli. The class $i)$/$ii$) is shown with
full/empty squares in all the panels of Fig. \ref{singlesbf}. The SBF
galaxies studied in C05 are also plotted in panels ($b$), with empty
circles, to enlarge the present sample with 6 giant ellipticals, and a
dwarf galaxy. All the galaxies of the C05 sample have a significant
I-band SBF versus color gradient, with the only exception of the local
dwarf NGC\,404 (shown with an arrow in panels $b$).

Inspecting Figure \ref{singlesbf}, we find that the SBF-gradient
galaxies tend to have fainter SBF magnitudes than SBF-flat galaxies,
with VCC\,941 being the only obvious exception to such behavior. More
specifically, the average SBF absolute magnitudes are $\langle \bar{V}
\rangle = -0.3 \pm 0.1$, and $\langle \bar{I} \rangle = -2.1\pm0.1$
for the SBF-flat objects, while they are $\langle \bar{V} \rangle = 0.7
\pm 0.2$, and $\langle \bar{I} \rangle = -1.5\pm0.3$ for the
SBF-gradient galaxies if the VCC\,941 data are excluded.  If also the
absolute SBF magnitudes for this latter galaxy are taken into account,
the differences between the SBF-flat and SBF-gradient galaxies are
less evident but still recognizable, as we have $\langle \bar{V}
\rangle= 0.6 \pm 0.5$ and $\langle \bar{I} \rangle =-1.5\pm0.3$.

Since the presence or absence of a SBF gradient is related to the
properties of the dominant stellar population, this feature could
represent a relevant tracer of galaxy formation. As an example, as
discussed in C05 (\S 4.2.1), the monolithic and hierarchical galaxy
formation scenarios make opposite predictions on the radial behavior
of the stellar population properties in a galaxy. In particular, in
the hierarchical scenario of galaxy formation the radial differences
of stellar population properties will flattens as galaxies undergo
mergers. On the contrary, substantial radial gradients are expected if
the galaxy formed following a pure monolithic collapse path
\citep[e.g.]{white80,bekki01}. 

In conclusion, the analysis of the SBF radial gradients in galaxies
might represent an interesting and innovative tool, to be used in
parallel with other techniques, to study the history of galaxy
assembly. SBF radial gradients are an observable physical quantity
that can be analyzed and compared between galaxies, independent of the
many model uncertainties.  However, in order to constrain galaxy
formation models using observed gradients, it is necessary to use
models to go from observational quantities to physical properties. As
will be shown in the next sections, currently the last step is very
uncertain in some metallicity regimes. In this respect, new
measurements are also necessary to enlarge the sample and to
strengthen the use of SBF gradients as a tool to analyze galaxy
formation.

As an additional way to examine such stellar population properties,
SBF and color data can be compared to population synthesis models. In
the following section we discuss the stellar population properties in
our sample of galaxies by comparing data with model predictions.

\subsubsection{Comparison with models and bias in the color transformations}

In this section we derive and discuss the properties of the
dominant stellar system in our sample of galaxies by comparing
observations of SBF and integrated color data with SSP model
predictions. At first we adopt the
Teramo-SPoT\footnote{http://www.oa-teramo.inaf.it/SPoT} models from
R05 as the reference ones. We adopt these models as reference since
they have proved to reproduce fairly well the Color--Magnitude
Diagrams, integrated magnitudes and colors, and the SBF amplitudes
both for star clusters (Galactic and Magellanic Clouds, MC, clusters),
and for galaxies in the optical and near--IR passbands
\citep[][R05]{brocato00,cantiello03}. We will also consider other sets
of models later on in this section.

In Figure \ref{single_sbf} we show the comparison of the average
($\bar{V}-\bar{I}$)$_0$ color and the absolute V- and I-band SBF
magnitudes versus the galaxy color. The R05 models are shown for ages
3$\leq$ t (Gyr) $\leq$ 14, and metallicity 0.0003 $\leq Z \leq$ 0.04.
In the Figure, models of equal metallicity are connected by dashed
lines.

Taking into account the upper two panels in this Figure, a good match
between models predictions and observational data is found, that
is, the observational data generally overlap the grid of models, with
the only exceptions of NGC\,2865 and NGC\,7626.

Even if the absolute magnitudes contain as additional uncertainty the
distance modulus adopted, the relevant result in these panels is the
good overlap between models and data, obtained for a wide range of
observed galaxy colors, i.e. from 0.85$\lsim (V-I)_{0} \lsim$ 1.30. In
this range, the observed $\bar{M}_I$ and $\bar{M}_V$ magnitudes
decrease by moving from blue to red integrated colors. According to
the R05 models shown in the figure, this means that the light emitted
by blue-faint galaxies is dominated by metal poor stellar populations
while red-bright galaxies are mostly populated by very metal rich
stars.

SBF color data can be used to derive the two color SBF versus
integrated diagram, which are independent of the distance modulus. By
inspecting the lower panel in Figure \ref{single_sbf} we find, as
expected, that the observed SBF and integrated colors cover a large
range of chemical compositions.  In this panel, it can be recognized
that the blue subpopulation, with an average $(\bar{V}-\bar{I})_0\sim
1.7 \pm 0.2$ lies in the area of models with Z$\lsim$0.001, possibly
higher than 0.0001. The red subpopulation, at $(\bar{V}-\bar{I})_0\sim
2.2 \pm 0.2$, seems more likely to be dominated by a
Z$\gsim$Z$_{\sun}$ stellar population. Moreover, while redder
galaxies, at (V-I)$_{0} > 1.2$, are mostly located near the edge of
old SSP models, the blue galaxies are spread over the whole age
interval.

The drawback with the $(\bar{V}-\bar{I})_0$ color is that the models
in the high metallicity regime are not well separated, leading to a
highly uncertain data to models comparison.  Note that such behavior
is predicted by all recent SBF models, as will be shown in the
following. In order to obtain more information from our SBF data, in
what follows we will consider the single annulus SBF and color
measurements for each galaxy, instead of the average values. This is
done in Figure \ref{spot} where we compare the annular absolute SBF
magnitudes and colors with R05 models. Again, the good matching of
models with data in the $\bar{M}$ versus (V-I)$_{0}$ planes is
encouraging, but we rather prefer to use the distance free SBF-color
versus color data and models comparison to infer the properties of the
stellar systems in the galaxy. Based on the content of the right panel
in Figure \ref{spot}, for each galaxy of our list we have obtained the
age and chemical composition reported in Table \ref{tab_ssp} (see also
the appendix for some more comments on individual galaxies).

First, we note that NGC\,2865, and NGC\,7626 data are noticeably far
from the grid of models. In general, it can be argued that our
measurements might suffer of the bias due to the ACS-to-standard
magnitudes transformations\footnote{It is worth noting that, by
inspecting the GC systems in NGC\,2865 and NGC\,7625,
\citet{sikkema06} have found that these galaxies show anomalous GC
$(V-I)_0$ color histograms.}. In fact, the observed SBF color range is
redder than the range used by \citet{sirianni05} to derive their
transformation equations. Furthermore, this bias could be stronger for
the measurements obtained from F606W images due to the uncertainty of
the F606W-to-V transformation -- e.g. Figure 21 from
\citet{sirianni05} shows that there is a non negligible difference
between the F606W-to-V synthetic and empirical transformations. As a
consequence, the V-band SBF data derived from F606W might be more
uncertain than others.

Although the presence of some bias in our data cannot be ruled out,
possible systematics can be highlighted by comparing these
measurements with those available in literature, derived in standard V
and I filters from ground based observations.

\citet{bva01} obtained $(\bar{V}-\bar{I})_0= 2.38 \pm 0.11$, for a
sample of galaxies in the Fornax cluster with M$_{Bt}\leq -20.3$
mag. For the galaxies presented in this work we find that the SBF
color for objects with M$_{Bt}\leq - 20.3$ mag is
$(\bar{V}-\bar{I})_0= 2.16 \pm 0.23$, but it becomes $2.29 \pm 0.08$
when NGC\,2865, and NGC\,7626 are excluded. If the latter number is
taken into account, we can consider our measurements and the ones from
\citeauthor{bva01} in good agreement. In addition, by comparing our
measurements for NGC\,1316 and NGC\,1344 with those from
\citeauthor{bva01} and T01, we find a good agreement for NGC\,1316,
while for NGC\,1344, that is one of the F606W-to-V SBF measurements,
the measurements still agree with each other, but to a lower degree.

The whole sample of objects with contemporary V and I SBF measurements
from the literature is shown in Figure \ref{mg2sig}, together with our
measurements. In this figure, the $(\bar{V}-\bar{I})_0$ data are
compared with the Mg$_2$ index and the central velocity dispersion
(both data from Leda). The GC data shown in the figure come from
\citet{at94}, and R05, the galaxies data are from
\citet{bva01}, \citet{tal90}, \citet{tonry90}, and this work. As
mentioned above, by inspecting this Figure we find that, if NGC\,2865,
and NGC\,7626 are excluded, the data presented in this work do not
show any peculiar behavior with respect to other (galaxies) data.

Moreover, comparing the left panel of Figure \ref{mg2sig} with the
Figures 14 and 15 from \citet{at94}, we find that the only ``unusual''
galaxy is NGC\,2865. More in detail, as discussed by \citet{at94},
there is a continuous trend in $(\bar{V}-\bar{I})_0$ versus Mg$_2$
from GC up to bright galaxies, but there seems to be a wide spread
when Mg$_2$ reaches $\gsim 0.3$. \citet{at94} argued that this could
be due to the behavior of the giant branch with metallicity, as the
``Mg$_2$ saturates before demonstrating the entire behavior of
$(\bar{V}-\bar{I})_0$''. In particular, the different saturation
limits of the blanketing with the metallicity in the V and I bands,
might cause the $\bar{V}$ to reach a saturation minimum brightness,
while the $\bar{I}$ still gets fainter at increasing metallicity. The
location of NGC\,7626 (and NGC\,4365, from the \citet{tal90} sample)
in the panels of Figure \ref{mg2sig} seems to confirm such behavior.

In other words, although our measurements could be affected at some
degree by the presence of a bias due to the ACS to standard magnitudes
transformations, the comparison of the present data with the data from
literature seems to exclude the presence of an overall bias
effect. Moreover, the unusual blue SBF color for NGC\,7626 could be
related to a differential metallicity saturation effect already
discussed by \citet{at94}. On the other hand, the behavior of
NGC\,2865 still appears peculiar with respect to the whole set of data
shown in Figure \ref{mg2sig}, possibly related to a bias in the data
(e.g. transformations), or to intrinsic properties of this galaxy (e.g.
dust, young stellar systems), or both\footnote{As a matter of fact
NGC\,2865 shows its peculiar behavior also when other, non-SBF, physical
properties are taken into account (Fig. \ref{single}).}.

A second consideration is that, using the R05 models, the model
predictions are generally confined to a narrow range of chemical
compositions for each galaxy, while the age limits are defined only in
few cases (e.g. VCC\,941). In the appendix, for each galaxy of
our list we quote the stellar properties derived using other
indicators; however, to further check the general age and metallicity
values obtained, here we compare our estimations with other
age/metallicity sensitive properties.

Concerning the chemical compositions, in Figure \ref{mgmbfeh} the
upper and lower bounds of the metallicity estimations drawn from the
models comparison are plotted against the metallicity sensitive index
$Mg_2$ (panel $a$), and with the total absolute magnitude M$_{Bt}$ of
the galaxy (panel $b$). From these panels we can recognize the
correlation of the Mg$_2$ index with metallicity, and the
metallicity-luminosity relation.  More specifically, it is confirmed
that the light from the blue sample of galaxies is dominated by a more
metal poor stellar component with respect to the red subsample.

We have additionally compared our age estimations with the age
sensitive index $H_{\beta}$. In this case, no significant
correlation emerges. However, it is worth noting that the $H_{\beta}$
measurements are confined to very small radii.

These findings should be considered as encouraging results in the
sense that they confirm the use of SBF magnitudes, color and gradients
as a valuable stellar population tracer mostly sensitive to the
chemical content of the galaxy.

One final comment, for the R05 models, is that the two-color diagram
in Figure \ref{spot} shows that the SBF color of red galaxies is not
well reproduced by models, with SBF color predictions systematically
$\sim 0.2$ mag redder than the observed ones. Moreover, one could
argue that the validity of stellar population properties presented in
this section is strictly model dependent. To encompass such evidence,
and test the robustness of predictions against models systematics, let
us we take into account several other recent SBF models derived from
SSP simulations.

In Figure \ref{others} we show the same comparison presented in Figure
\ref{spot}, but for the \citet{bva01}, \citet{liu02}, and
\citet{marin06} (BVA01, L02, and MA06 respectively hereafter) stellar
population models.

As a first comment, we notice that at the high metallicity regime
(Z$\gsim$0.01) the SBF magnitudes versus (V-I)$_0$ color behavior is
quite similar for all models. This is probably due also to the fact
that the topology of the grid of models here can be very complex,
leading to high degeneracy. On the other side the slight mismatch for
the red galaxies noted with the R05 models is still present in these
other sets of models. As suggested by BVA01, this could be due to the
use of SSP models, while composite stellar populations would be more
appropriate. Additionally, we mention that in most of the red galaxies
SBF gradients (i.e. stellar populations gradients) have been
found. Moreover, it must be noted the non--negligible differences
between the various models in the low-Z regime.

Keeping in mind such limits of SSP models, we have analyzed the
location of each galaxy in the SBF-color versus color plane with
respect to models in order to derive the stellar population
properties, as for the R05 models. The results are reported in
Col. 3-5 of Table \ref{tab_ssp} (see also appendix). As can be
recognized from the data in the Table, the chemical composition ranges
obtained with the different models show a general good overlap. 

On the other hand, we find that the ranges of acceptable ages are
quite large, similarly to what obtained with R05 models. Thus, a
robust conclusion of this work is that SBF can provide, given the
current stage of the models, reliable estimates of the typical
metallicities of galaxies, while the age of the dominant stellar
system cannot be well constrained with this technique.

The right panel of Figure \ref{mgmbfeh} shows the metallicity versus
Mg$_2$ and M$_{Bt}$ comparisons already discussed for the R05 models,
except that in this case the average metallicity from the four
different models is considered. The [Fe/H]--Mg$_2$ correlation is
still present, although the data refer only to the red sample of
galaxies. The relationship between [Fe/H] and M$_{Bt}$ is more
interesting. In fact, a mean least--squares fit to the data shown in
Figure \ref{mgmbfeh} (panel $d$) yields: [Fe/H]$\sim$ (-0.12 $\pm$
0.02) $\times$ M$_{Bt}$ - (2.8 $\pm$ 0.3). The agreement of this
equation with similar relations existing in literature
\citep[e.g.][]{kobulnicky99,contini02} leads us to conclude that,
within the limits of the present treatment, the method proposed
provides us with reliable ranges of acceptable metallicities.
Therefore, the metallicity properties derived from models can be
considered by and large acceptable. Once again, no noteworthy
correlation of ages with the H$_{\beta}$ index is recognizable.

Let us now look more in detail to the model predictions for the single
galaxies. Doing such comparison, consistent results with different
models would imply a significant constraint on the properties of the
stellar system observed, on the other hand a lack of agreement between
models will possibly highlight uncertainties in the theoretical
predictions.

Inspecting single galaxies it can be seen that at high metallicity
(Z$\gsim$ 0.01) the differences between model predictions are less
severe with respect to the case of Z$\lsim$0.001. More in detail, the L02
and R05 models predict that age variations in the low metallicity
regime do not affect the SBF color, while integrated colors suffer a
noticeable change. On the contrary, BVA01 and MA06 models predict
almost constant integrated colors at different ages, and a substantial
SBF color variation.

The differences at low metallicity are possibly originated by a
different treatment of the evolutionary properties of the bright, cold
stellar component along the RGB and AGB in the various models.  Until
such discrepancies are resolved, little can be said about the origins
of SBF gradients in this metallicity regime. However, such differences
are interesting for the purpose of refining the models for the
evolution of RGB and AGB stars. As well known, in fact, the SBF
magnitude at these wavelengths is very sensitive to the properties of
RGB/AGB stars, which suffer from large uncertainties, e.g. in
atmosphere models, mass loss and stellar wind. In the optical regime,
at lower metallicities the giant branch is brighter with respect to
higher metallicities, thus it has a stronger effect on the luminosity
weighted SBF signal. As a consequence, the discrepancies between
various models in the treatment of the bright RGB/AGB phases appear
more clearly in the SBF models at the low Z regime.  Adopting this
view, the disagreement between models can be solved by refining the
modeling of the evolutionary properties of giant branch stars. Or,
viceversa, coupling the bright star sensitive SBF color and integrated
color data with metallicity and age information from independent
indicators, would provide information useful to the challenge of a
refined modeling of RGB/AGB stars.

\subsubsection{SBF gradients and models}

In addition to the above considerations, we now discuss the gradients
of stellar populations properties, also taking advantage of some C05
results. As noted before, differently from C05 where we succeeded in
revealing the systematic presence of SBF gradients in seven elliptical
galaxies over eight, in our new sample of data we do not find a
systematic presence of SBF gradients.

Before going on with the discussion of the gradients we must emphasize
that such discussion, as for the results presented in the previous
sections, is highly model--dependent. However, in contrast to the
previous comparisons, where we compared observations with absolute
values of SBF and colors, here we are only considering the slopes of
such SBF versus color relations. On the other hand, it is easy to
recognize that the slopes of the SBF versus (V-I)$_0$ relations can
change quite a lot depending on the models. For example, if one
considers the slope $\delta \bar{M}_I / \delta (V-I)_0$ at fixed Z, it
has an average value of 3.3 for the R05 models, but it can be as high
as $\delta \bar{M}_I / \delta (V-I)_0 = 14.3$ if the BVA01 models are
considered. As explained before this is mostly due to the strong
differences between models at the low metallicity regime. If only
models at $Z\gsim 0.01$ are considered, in fact, such differences
disappear, as we find that $\delta \bar{M}_I / \delta (V-I)_0$ at
fixed metallicity lies in the range of 3.2-4.5 for all the models. At
the same time the $\delta \bar{M}_I / \delta (V-I)_0$ at fixed age
lies in the range 7.3-9.7, in good agreement for all the models at
$Z\gsim 0.01$.

These numbers tell us that any obvious SBF-gradient in the low
metallicity regime cannot be interpreted clearly as age--driven or
metallicity--driven gradients, because of the differences between
models. On the contrary, in the high-Z regime all models predict that
a gradient related to pure age variations has a SBF versus color slope
one half the slope of a gradient due to pure age
variations. Obviously, real galaxies do not follow the simple scheme
of ``pure age(metallicity) radial variations'', however, as done in
the previous sections, our observational data can be used to set some
constraints at least to the effective dominant stellar system in the
galaxy. Keeping in mind these limits, let us make some considerations
on the SBF-gradients for the galaxies located in the area of high-Z
models.

In C05, we have found that the gradients were mainly explained by
metallicity variations within the galaxy, the inner regions being more
metal rich than the outer ones. In the new sample we find that in the
cases where a stellar population gradient is evident (SBF-gradient
galaxies in Table \ref{tab_measures}), it is mostly explained by a age
gradient rather than by a metallicity one. For example, this is the
case of NGC\,1344, where $\alpha \equiv \delta \bar{m}_I / \delta
(V-I)_0 =2.8 \pm 0.3$. However, in few cases (e.g., UGC\,7369 where
$\alpha = 9.0 \pm 3.2$) a metallicity gradient is probably observed,
no matter what set of models is considered.

For the galaxies in the low-Z regime, as already mentioned, the
diversity between models hampers any clear justification of the source
of the gradient itself, and different models predict opposite
explanations to the presence of the gradient.

The differences between the present results and those by C05 can be
ascribed to the sample selection. In fact, the list of objects taken
into account in this work covers a wider range galaxy properties with
respect to the C05 sample, which is mostly limited to massive
ellipticals.  In C05, the only exceptions to the uniform SBF gradient
behavior were NGC\,1344 and NGC\,404, that is the galaxies with the
lowest mass in the sample. For NGC\,1344 an age--driven gradient was
predicted, as also confirmed by the SBF color measurements in this
work; while for the local dwarf galaxy NGC\,404 no sign of evident age
or chemical composition gradient was recognized, as it is the case for
almost all the present dwarf galaxies.

By coupling the data in this work, with the results from C05 we are
lead to the general indication that normal/bright ellipticals
preferentially show a metallicity gradient, with the inner galactic
regions being more metal rich with respect to the outer ones.  

On the other hand, we have found that in some galaxies it is very
likely that the observed gradient is probably due to a radial change
of age, although the absolute age estimation is not feasible using
available models. Most of such galaxies, like NGC\,1344, show evidence
of morphological irregularities indicative of recent merging activity.

Finally, the less massive objects invariably show no evident signs of
SBF-gradients, or, as in the case of VCC\,941, the gradient
cannot be obviously interpreted in terms of metallicity or age
variation effects, because of the disagreement existing between
different models in this metallicity regime.

In conclusion, our SBF color versus integrated color study seems to
point out that there is a metal enrichment in the stellar populations
at inner galaxy regions, and that such behavior is related to the
galaxy mass, as it can only be recognized in the more massive
objects. The possible age--driven gradients, also observed in
our sample, are mostly related to specific environmental properties
(e.g. galaxy interactions).

Before closing this section, we say a few words about the consequences
of the metallicity properties derived above on the RGB Tip
distances. All the galaxies of the blue subsample, in fact, come from
proposals designed to derive RGB Tip distances of the target
galaxy. By using the R05 models we find that for all the galaxies at
(V-I)$_0\lsim 1$, the age and chemical compositions limits given above
imply a RGB Tip magnitude in the I-band which is practically constant
for the whole intervals quoted, i.e. $M_{I,RGBTip}\sim -4.2$ mag. The
model predictions are slightly brighter with respect to the
observational RGB Tip magnitudes adopted by \citet{karachentsev06},
which use $M_{I,RGBTip}\sim -4.05$ mag. However, we must mention that
$i$) both the theoretical and observational data agree within
uncertainties; $ii$) the light of galaxies at (V-I)$_{0}\lsim 1$ is
expected to be dominated by a metal poor (Z$\lsim$0.001) stellar
system, thus brighter RGB Tip magnitudes are expected
\citep[e.g.][]{salaris98}; and that $iii$) the theoretical predictions
do fully agree with the recent calibration from
\citet{rizzi07}. Adopting the above theoretical calibration implies an
average of 8\% higher distances with respect to the one adopted by
\citet{karachentsev06}.

\subsection{Determining the best $\bar{M}$ versus (V-I)$_0$ calibration}
Since the first appearance of the SBF method, a few different
calibrations of the $\bar{M}_I$ absolute SBF magnitude versus the
(V-I)$_0$ color have been introduced, by using either observations, or
theoretical models. In this section we focus our attention on
empirical calibrations reviewing the most recent ones \footnote{ We
exclude from this section the data of NGC\,2865 due to the peculiar
behavior of this galaxy (\S 4.1.2, and Appendix).}. Typically such
equations are derived from different observational data, and they are
valid only within the range of colors of the defining sample, which is
basically narrower than the color range of the present sample.  We
apply these calibrations to our set of measurements and compare the
distance moduli so derived $\mu_{0,cal}$ with the group ones
$\mu_{0,group}$ obtained from literature data (Table
\ref{tab_dati}). The minimization of the $\Delta \mu_0 =
\mu_{0,cal}-\mu_{0,group}$ versus the calibration equations, will
enable us to identify the best empirical calibration in the color
interval 0.85$\lsim (V-I)_{0} \lsim$ 1.30.  For this study we use the
average SBF and color measurements from Table \ref{tab_measures}.

Among the few available, we will consider the following calibrations:

\begin{equation}
\bar{M}_{I,T01} = -1.74 \pm 0.08 + (4.5 \pm 0.25) [(V-I)_0 -1.15]
\label{eqt01}
\end{equation}
from T01, obtained using a sample of $\sim$ 300 galaxies
in different groups, zeropoint magnitude calibrated using the $HST$
Key Project Cepheids distances by \citet{ferrarese00}.

Then we consider the other:
\begin{equation}
\bar{M}_{I,J03} = -1.58 \pm 0.08 + (4.5 \pm 0.25) [(V-I)_0 -1.15]
\label{eqj03}
\end{equation}
from \citet{jensen03}, which is essentially the same as the
Eq. \ref{eqt01}, but the zeropoint is derived using the revised $HST$
Key Project Cepheid distances from \citet{freedman01}, without the
metallicity correction. Both equations are valid in the range of color
0.95 $\leq$ (V-I)$_0 \leq$ 1.30.

In combination with these equations, for those objects at (V-I)$_0
\leq 1.00$ we apply the calibration derived by \citet{at94} from
globular clusters, upgraded by using the most recent distances and
extinctions from the \citet{harris96} catalog\footnote{Available at
the web address http://www.physics.mcmaster.ca/$\sim$harris/mwgc.dat},
including also the measurements for old (t$\gsim$ 10 Gyr) LMC globular
clusters from R05. With all these upgrades we obtain:
\begin{equation}
\bar{M}_I = 2.17 \pm 0.27
\label{eqggc}
\end{equation}
using the star clusters with (V-I)$_0 \leq 1.00$.

Finally, we also take into account the recent calibration from
\citet{mieske06}, derived from Fornax cluster galaxies, in the color
range 0.85 $\leq$ (V-I)$_0 \leq$ 1.10:

\begin{equation}
\bar{M}_{I,M06} = -2.13 \pm 0.17 + (2.44 \pm 1.94) [(V-I)_0 -1.00],
\label{eqm06}
\end{equation}

we coupled this equation with Eq. \ref{eqj03} for galaxies at
(V-I)$_0 > $ 1.10.

The calibration from \citet{ferrarese00} is not considered here as it
agrees within uncertainties with Eq. \ref{eqt01}. These authors, in
fact, using a similar approach to T01, adopted the same slope of
Eq. \ref{eqt01}, but found a zero point of -1.79 $\pm$ 0.09 mag.

In Table \ref{tab_distances} we show the distance moduli obtained
adopting the above calibrations. In the Table we also show the group
distance modulus for each object, which is used to derive the reduced
$\chi^2$, and the average differences $\Delta \mu_0 \equiv \langle
\mu_{0,cal} -\mu_{0,group} \rangle$. Both the quantities $\chi^2$ and
$\Delta \mu_0$ are reported in the last rows of the Table.

As shown by the numbers in the last two rows of Table
\ref{tab_distances}, the best matching with the group distances is
obtained coupling equation \ref{eqj03} with the \ref{eqggc}, namely:

\begin{eqnarray}
\bar{M}_I = -1.58 \pm 0.08 + (4.5 \pm 0.25)\times [(V-I)_0 -1.15],~~~~1.00 < (V-I)_0 \leq 1.30  \label{eqbest1}
\\
\bar{M}_I = -2.17 \pm 0.27,~~~~0.80 < (V-I)_0 \leq 1.00. 
\label{eqbest2}
\end{eqnarray}

We note that if we apply the latter equations to the SBF measurements
for 25 Fornax Cluster galaxies from \citet{mieske06}, a median
distance modulus 31.3 $\pm$ 0.4 is obtained, which is consistent with
the expected group distance for this cluster $\mu_0 \sim$31.5.

As an additional check, we have carried out for our V-band
measurements the same analysis performed on I-band calibrations. The
main difference in this case is that there is one only recent
empirical calibration, from BVA01. These authors provide:
\begin{equation}
\bar{M}_{V,BVA01} = 0.81 \pm 0.12 + (5.3 \pm 0.8) [(V-I)_0 -1.15].
\label{eqv}
\end{equation}

Although this equation has been derived using data in a narrow range
of colors (1.05$\leq (V-I)_0 \leq$ 1.25), we tentatively extend its
validity to the same range of Eq. (\ref{eqbest1}) colors -- such
assumption is not completely arbitrary: in fact, as shown by numerical
simulations, at fixed age the V-band SBF magnitudes have a more linear
behavior versus $(V-I)_0$ respect to the I-band \citep[e.g., Figures
\ref{spot}-\ref{others}, and Figure 5 in][]{cantiello03}. We have
renormalized the BVA01 zeropoint using the same criteria adopted by
\citet{jensen03}.

Again, for the blue galaxies we adopt the calibration derived from
globular clusters using the \citet{at94} and R05 measurements. In
this case we obtain:
\begin{equation}
\bar{M}_V = -0.50 \pm 0.27.
\label{eqggcv}
\end{equation}

Coupling the latter two equations with our SBF and color measurements,
we obtained the distance moduli also reported in Table
\ref{tab_distances} (Col. 7).

Using the best I- and V-band calibrations
(Eqs. \ref{eqbest1}-\ref{eqbest2}, and \ref{eqv}-\ref{eqggcv},
respectively) we obtain the weighted average distance moduli reported
in the last column Table \ref{tab_distances}. Once more, we note that
the general validity of the calibrations is shown by the satisfactory
agreement between the distance moduli derived and the group distance
moduli.

As an aside, we also derived an independent calibration for the
$\bar{M}_V$ versus (V-I)$_0$ equation, coupling our measurements with
other data from literature. As a result we have found that the
calibration obtained agrees within uncertainties with
Eq. \ref{eqv}-\ref{eqggcv} in the whole range of $(V-I)_0$ colors
considered here.

Before concluding this section we point out three facts.  First,
\citet{karachentsev06} obtained a distance modulus $\mu_0 = 30.32$ for
UGC\,7369, based on the RGB Tip method.  However, they also state that
this galaxy ``does not look to be a nearby object'', and that it is
``plausible association with the Coma I group'' at a distance modulus
31.07 $\pm$ 0.07 (T01). As shown by the data in Table
\ref{tab_distances} our SBF measurements support this last hypothesis.

Second, it has been widely discussed by \citet{richtler03} that the
Globular Cluster Luminosity Function (GCLF) is a quite reliable
distance indicator, although some exceptions exist. In his review,
\citeauthor{richtler03} uses the SBF distances from T01 to derive the
GCLF absolute Turn Over Magnitude (TOM). One of the main exceptions to
the universality of the GCLF is NGC\,3610, whose TOM is $\sim$ 2
magnitudes fainter than expected. \citeauthor{richtler03} argues that
one of the possible causes of such mismatch is the presence of a
population of intermediate-age metal-rich clusters, resulting in a
fainter TOM. However, even though only the blue subpopulation of
clusters is taken into account, there is still a large offset between
NGC\,3610 and the other galaxies. We find worth noting that, in
their recent study on the GC system of NGC\,3610, \citet{goudfrooij07}
have found $M_V^{TOM} \sim -7.2$ mag. However, these authors
erroneously state that they adopt a ``distance modulus of
$(m-M)_0=32.65$ as measured from T01''. The distance modulus quoted
for this galaxy by T01, in fact, is 31.65 $\pm$ 0.22, which leads to
$M_V^{TOM} \sim -6.2$, one magnitude fainter than the typical value
for normal elliptical galaxies.

On the other hand, if we adopt the average distance modulus of
NGC\,3610 from Table \ref{tab_distances}, $\mu_{Ave} = 32.71 \pm
0.08$, and the blue clusters TOM 25.44$\pm$0.10 from
\citet{whitmore02}, the absolute TOM is $M_V^{TOM}=-7.27 \pm 0.13$, in
good agreement with the universal TOM from \citeauthor{richtler03}
$M_V^{TOM}=-7.35 \pm 0.24$\footnote{We have corrected the mean
$M_V^{TOM}$ from \citeauthor{richtler03} value applying the -0.16 mag
zeropoint shift as discussed in \citet{jensen03}.}. The
difference between our distance and the estimation from T01 is
probably due to the much lower quality of the NGC\,3610 ground-based
data used by T01, compared to these high resolution ACS
images. Inspecting the data quality flags from T01 (see their Table 1,
Q and PD values), we find that the SBF magnitude of NGC\,3610 should
be considered as poorly constrained. Moreover, the T01 distance of the
other galaxy NGC\,3613, which is classified as same group member of
NGC\,3610, agrees within uncertainties with our new SBF distance.

Finally, by using the average distance moduli in the last column of
Table \ref{tab_distances}, and the $v_{flow}$ values reported in Table
\ref{tab_dati}, we obtain $H_0 \sim 71 \pm 14$ km s$^{-1}$
Mpc$^{-1}$, if the galaxies with $v_{flow} \gsim~1000~km/s $ are taken
into account.

\section{Conclusions}

The SBF and color properties obtained from ACS V- and I-band images of
14 galaxies have been discussed. The data were drawn from the HST
archive. Classical integrated and SBF magnitudes have been derived
using the standard analysis procedures. Our set of measurements is
unique in terms of the wide range of $(V-I)_0$ color. We have taken
advantage of this property to address different questions concerning
both the use of SBF measurements as a distance indicator, and as a
stellar population tracer.

With regard to the use of SBF to study stellar populations issues, we
have analyzed the properties of the dominant stellar population in the
selected galaxies by coupling V- and I-band SBF magnitudes with the
$(V-I)_0$ color of the galaxies. As expected the list of objects
covers a wide range of stellar populations properties. Since the
outcome of this study depends on the set of population synthesis
models adopted, we have taken into account different models to test
the robustness of the predictions against the models systematics.
Different sets of SSP models typically provide chemical composition
estimations similar to each other and to our reference R05
models. However, from this comparison we have found that generally
{\it it is not possible to derive reliable age constraints} for the
stellar component in each galaxy, due to the non--negligible
differences between models, especially at lower metallicities. In
other words, multi--models comparison has shown that this technique is
not efficient to strongly constrain the age for old (t$\gsim$3 Gyr)
stellar systems, but it can realistically be used to confine the
metallicity range of the stellar system that dominates the light
emitted by the galaxy.

These results confirm the usefulness of this kind of analysis to
investigate the evolutionary properties of the unresolved stellar
component in distant galaxies. On the other hand -- where model
differences arise -- they also indicate that the present knowledge of
stellar evolution, with particular regard to the properties of cool,
bright giant branch stars, is still an open question which could be
efficiently challenged taking advantage of SBF color data.
 
We have also examined radial SBF behavior for the sample of 14
galaxies. Comparing the gradients with models, and taking also into
account the results from C05, we observe that usually the dwarf
galaxies do not show substantial SBF gradients, thus we do not
find any sign of systematic radial age/metallicity
variation. Moreover, where such gradients are observed (e.g. VCC\,941)
the opposite predictions made by different sets of models make it
difficult to understand if such gradients are related to radial
changes of age or metallicity.  On the contrary, for more massive
objects a preferential metallicity driven gradient is noticed, with
the outer galaxy regions being more metal poor than the inner
ones. Possible age gradients have also been found, however they
are usually related to a recent merging event. As a consequence, our
SBF gradient data seem to point out the existence of a mass--related
metallicity gradient in spheroidal galaxies. Given the connection
between the gradients of stellar populations properties (i.e. SBF- and
color-gradients) and the possible galaxy formation scenarios, we
suggest that a future, enlarged database of SBF gradient data will
also provide a valuable tool to trace the history of galaxy formation.

In conclusion, our study illustrates the potential of a study of
galaxy properties based on the comparison of SBF colors with
populations synthesis predictions. The current state of SBF
models allows for a robust determination of the mean metallicities of
galaxies, and an improved understanding of the stellar evolution
phases important for SBF might allow the use of SBF in the future for
detailed population studies. As in \citet{cantiello03}, again we
remark that multi--wavelength SBF data involving optical to near--IR
observations are of paramount interest to push forward this
technique. As shown by model predictions, indeed, SBF colors like
$\bar{B}-\bar{K}$ are not affected by the models degeneracy shown by
the $\bar{V}-\bar{I}$ color. Additionally, such color data are
sensitive to stars in different phases of their evolution -- e.g.,
$\bar{B}$ to Horizontal Branch stars, $\bar{K}$ to Thermally--Pulsing
AGB stars -- and are expected to be much more efficient to trace the
stellar content of the galaxy, that is to trace back the history of
galaxy formation.

Concerning distance studies, to check the validity of some I-band (and
V-band) empirical calibrations existing in literature, we have
estimated the galaxy distance moduli coupling our data with the
various empirical calibrations. Then, these distance moduli have been
compared with group distances derived from literature.  We have found
that the best I-band calibration is obtained matching two relations:
$a$) in the range 1.00$<(V-I)_0\leq$1.30 the equation by
\citet{jensen03}, which is basically the one obtained by T01 with a
different (fainter) zeropoint; $b$) for the range of color
0.80$\leq(V-I)_0\leq$1.00, a constant absolute magnitude, derived from
Galactic and MC globular clusters. Adopting a similar approach with
the V-band data, we have found that the calibration provided by BVA01
extended to interval 1.00$<(V-I)_0\leq$1.30, with a constant SBF
magnitude for colors within 0.80$\leq(V-I)_0\leq$1.00, gives SBF
distances in good agreement with group distances.

Using the best I- and V-band calibrations, and taking into account
only the galaxies at $v_{flow}\geq$1000 km/s, we estimated $H_0 \sim
71 \pm$ 14 Km s$^{-1}$ Mpc.

\acknowledgments

This work was supported by the NASA grant AR-10642, by COFIN 2004
under the scientific project ``Stellar Evolution'' (P.I.: Massimo
Capaccioli), and by PRIN-INAF2006 "From local to cosmological
distances" (P.I. G.Clementini).

\appendix
\section{A comparison of the observed $\bar{V}-\bar{I}$ colors with models. Comments on
individual galaxies}

Based on the content of the right panel in Figure \ref{spot}, and
Figure \ref{others}, in the following we discuss the chemical and
physical properties each single galaxy of our sample by
interpolating between models at different ages and chemical
compositions.

\begin{figure}[!ht]
\plotone{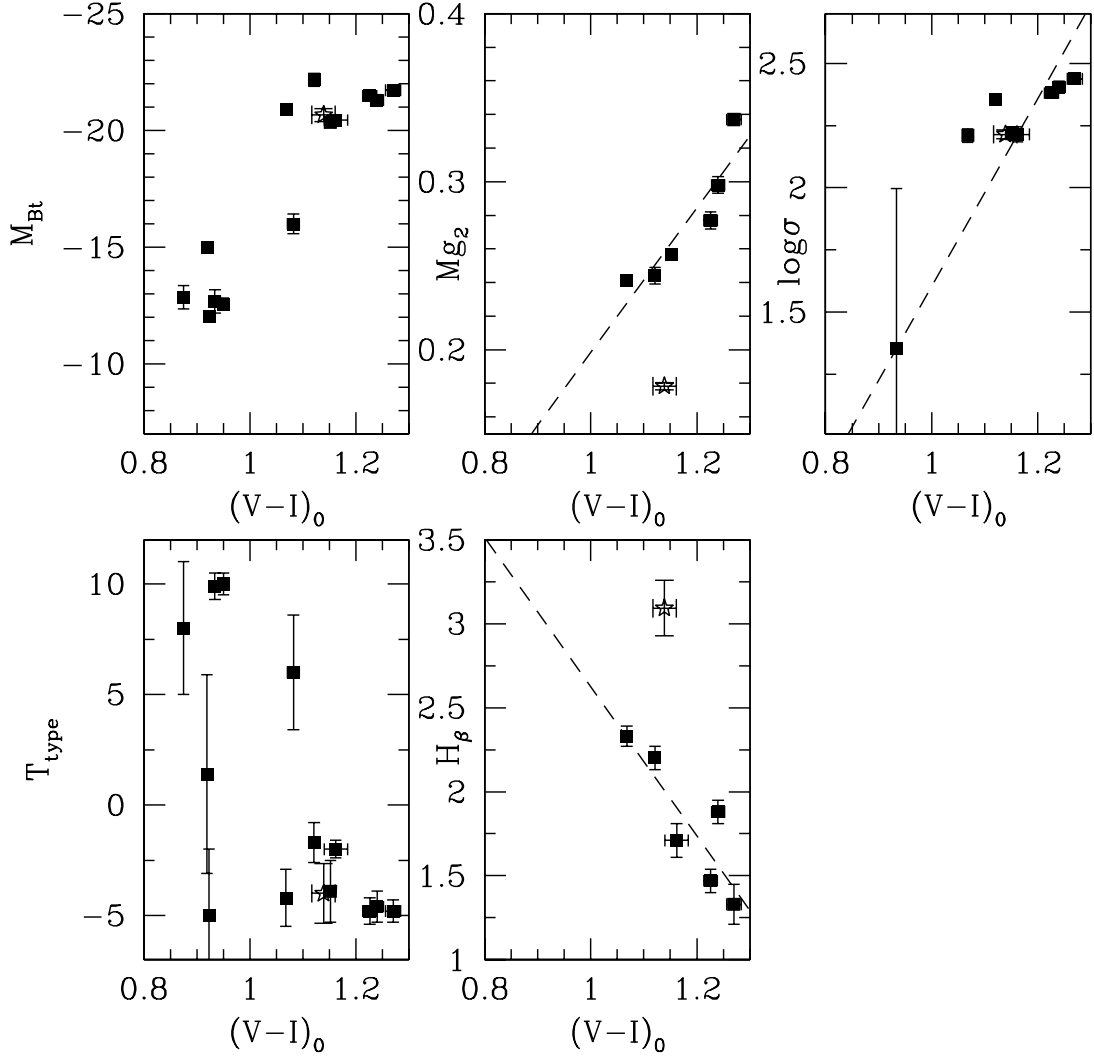}
\caption{Observational properties of the sample of galaxies versus the
integrated (V-I)$_0$ color. NGC\,2865 data are marked with a
five-pointed star.The least--squares dashed lines are obtained
excluding NGC\,2865 from the fit.
\label{single}}
\end{figure}

\begin{figure}[!ht]
\epsscale{1.1}
\plotone{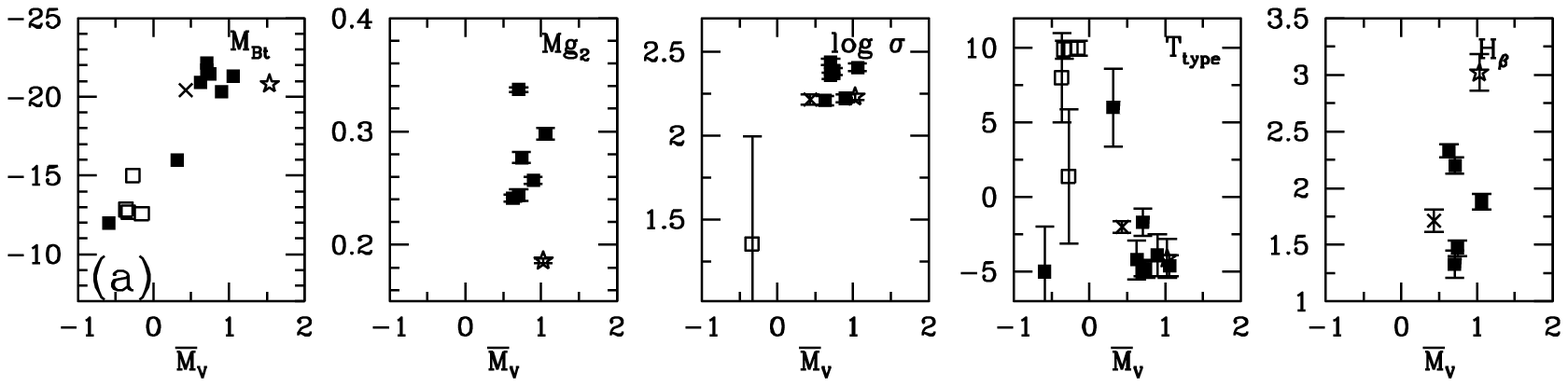}
\plotone{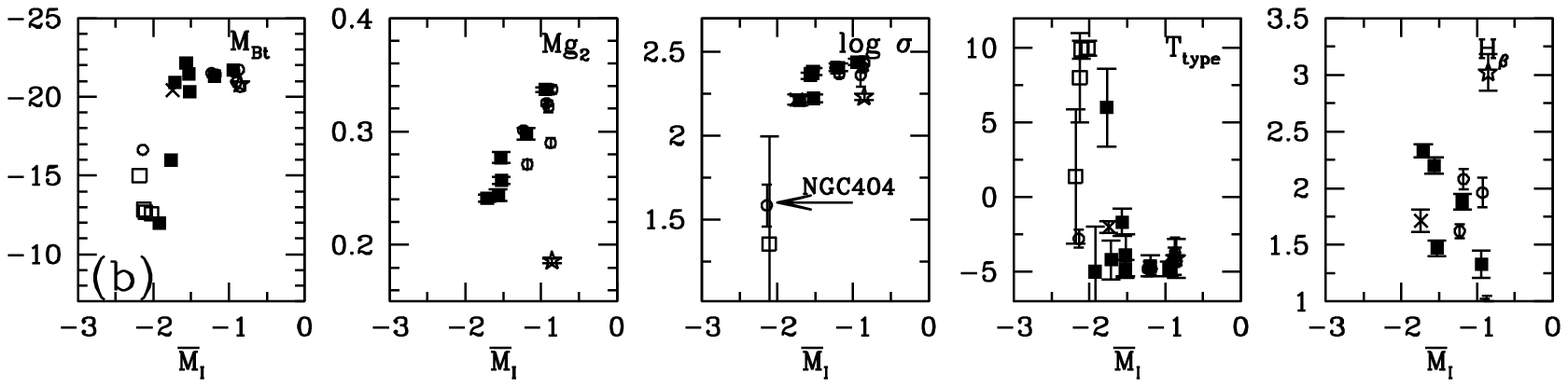}
\plotone{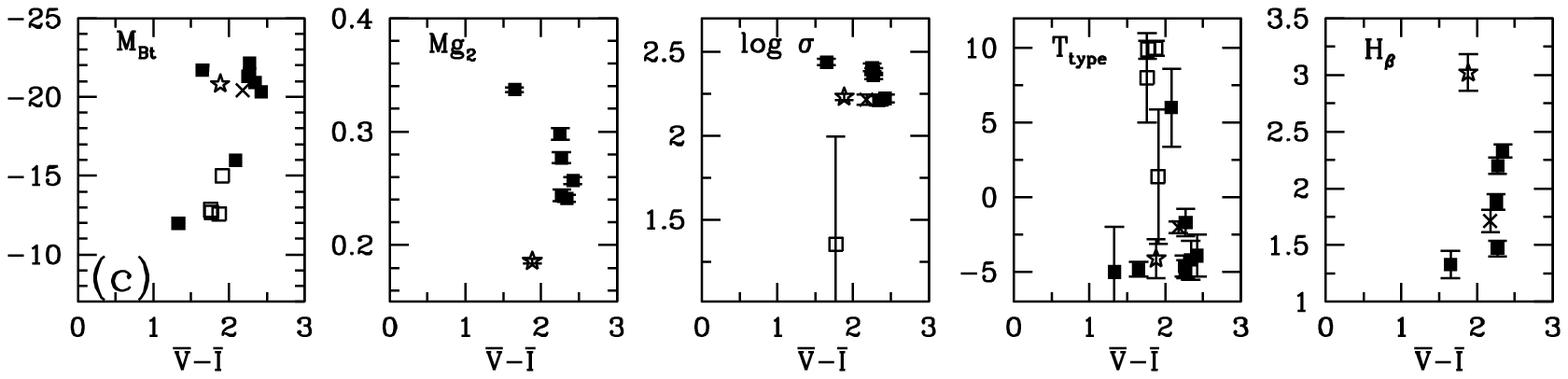}
\caption{The same observational properties shown in Figure
\ref{single} (upper quote in each panel) plotted against the absolute
V-band SBF magnitudes (panels $a$), I-band absolute SBF (panels $b$),
and the SBF $(\bar{V}-\bar{I})_0$ color (panels $c$). In all panels
empty/full squares show the location of galaxies without/with a
significant SBF gradient.  NGC\,2865 and NGC\,474 data are marked with a
five-points star and cross, respectively.  In addition to the data in
Table \ref{tab_measures}, the C05 galaxies data are also shown with
empty circles in panels $b$.  All the galaxies from the C05 sample have
a substantial gradient, except the dwarf NGC\,404.
\label{singlesbf}}
\end{figure}

\begin{figure}[!ht]
\epsscale{1.}
\plotone{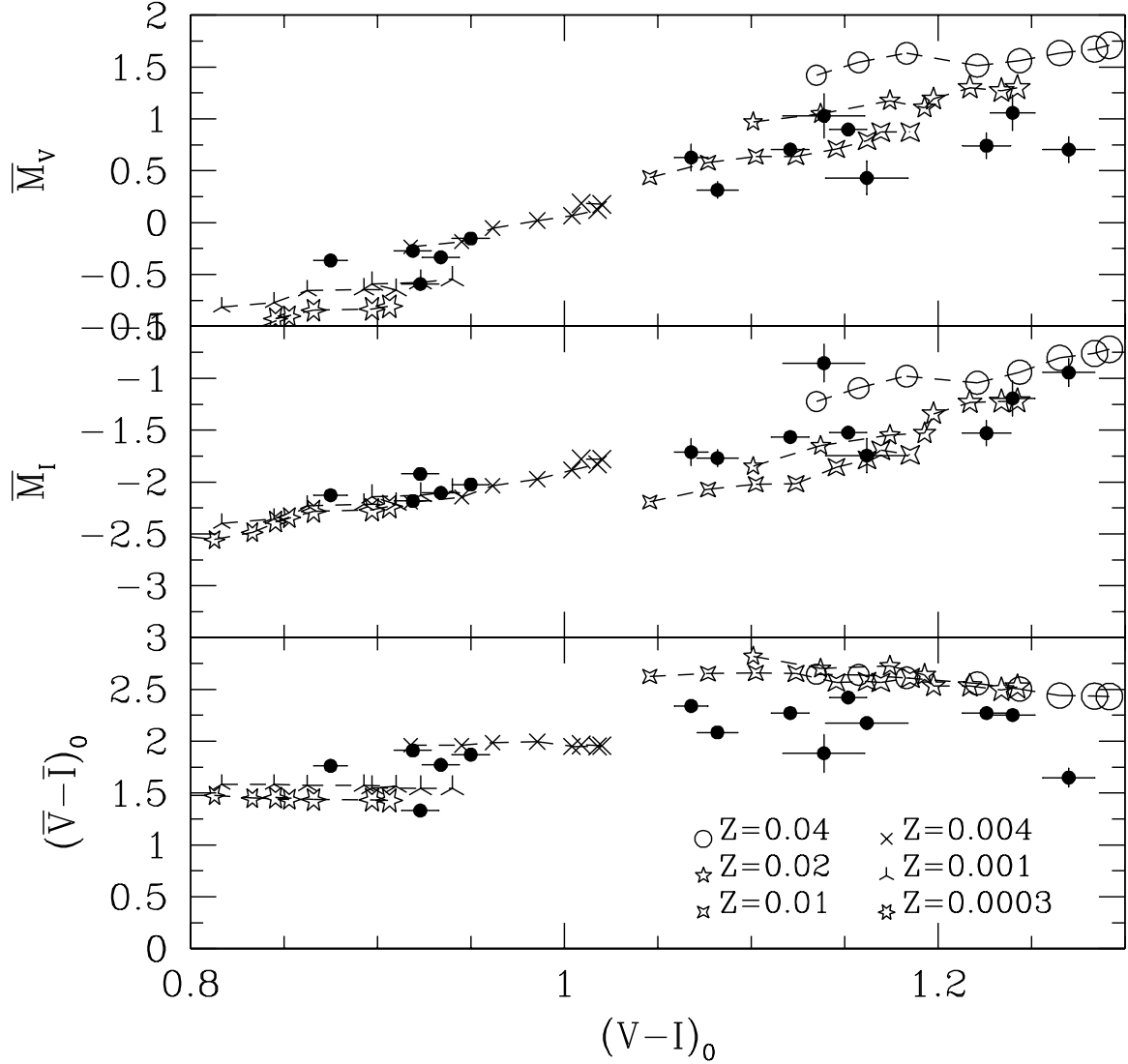}
\caption{SBF magnitudes versus the integrated color. The average data
from Table \ref{tab_measures} are plotted. The R05 models for
t=3, 4, 5, 7, 9, 11, 13, 14 Gyr are over-plotted in the panels. Increasing
symbols size refer to older ages. Different symbols shape mark
different chemical compositions, as labeled.
\label{single_sbf}}
\end{figure}

\begin{figure}[!ht]
\epsscale{1}
\plottwo{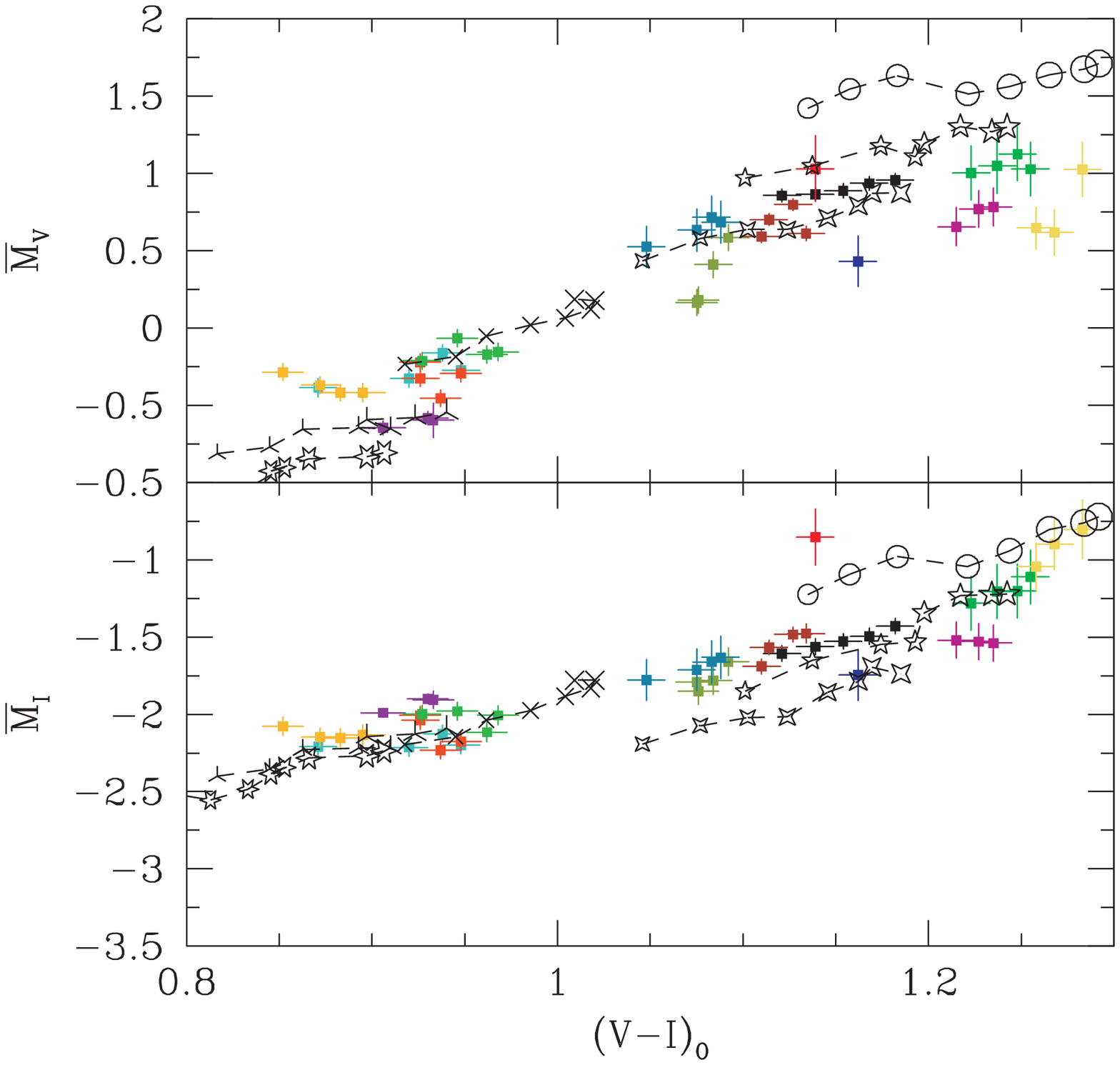}{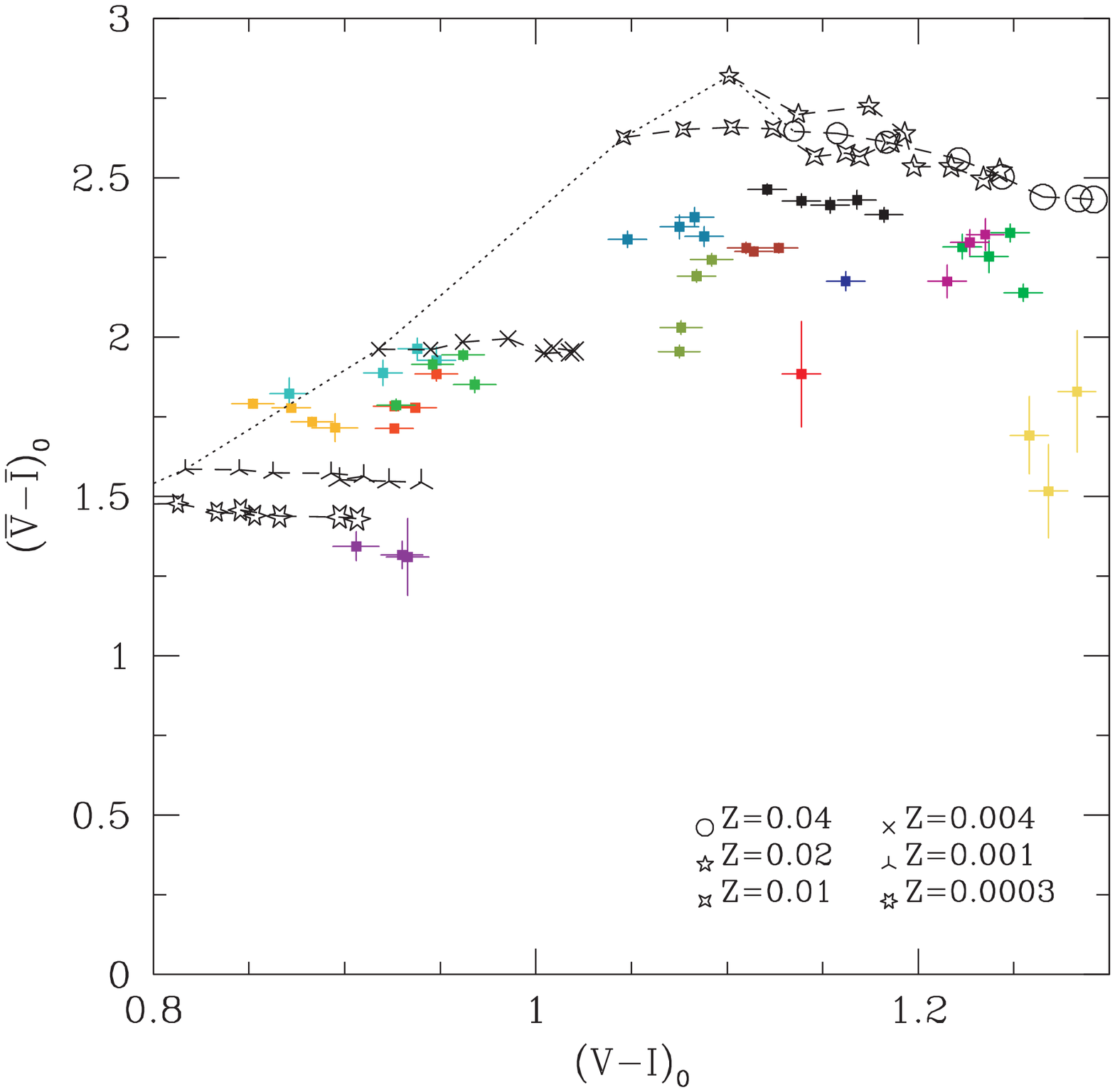}
\caption{Left panels: Absolute SBF profiles versus the integrated
(V-I)$_0$ color. Right panel: The distance free SBF-color versus
integrated color panel.  R05 models symbols are the same as in Figure
\ref{single_sbf}. The dotted line connects the models at lowest
ages, 3 Gyr, for all chemical compositions.
\label{spot}}
\end{figure}

\begin{figure}[!ht]
\plottwo{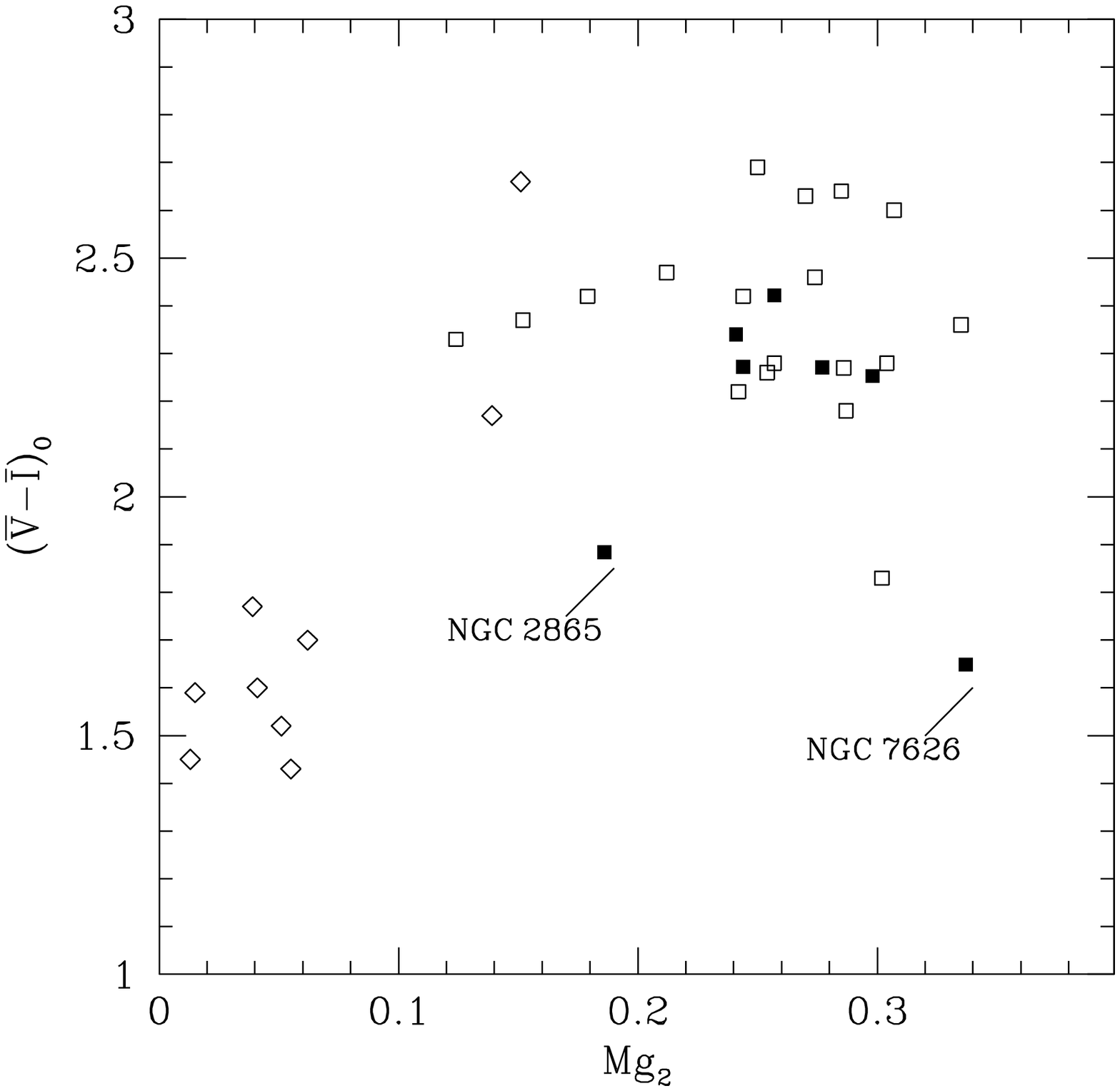}{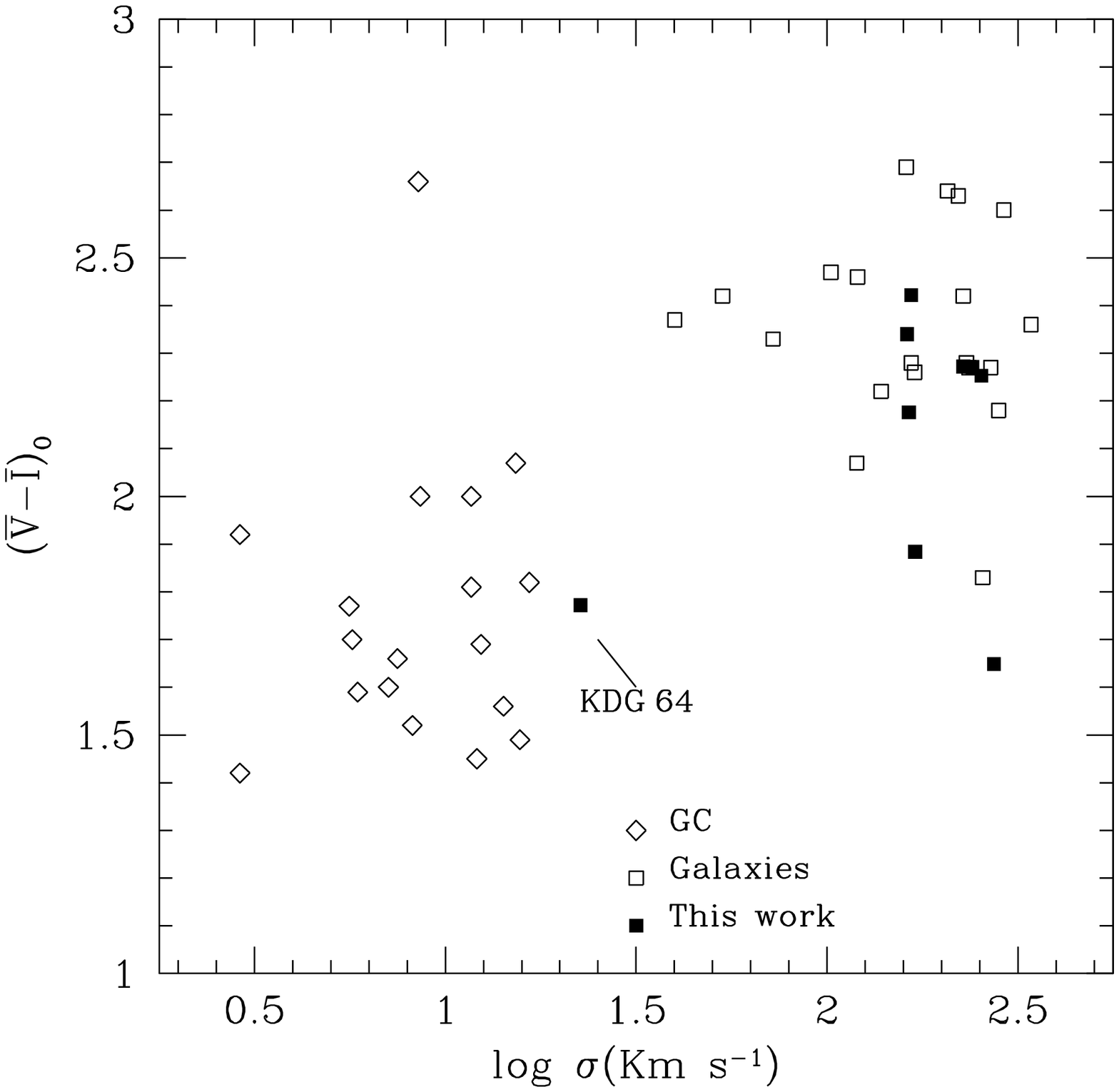}
\caption{SBF  $(\bar{V}-\bar{I})_0$ color versus the 
Mg$_2$ index (left panel), and versus the central velocity dispersion
(right panel). Empty symbols show data taken from literature: diamonds for GC,
squares for galaxies. Full squares refer to the measurements from this work. 
\label{mg2sig}}
\end{figure}

\begin{figure}[!ht]
\plottwo{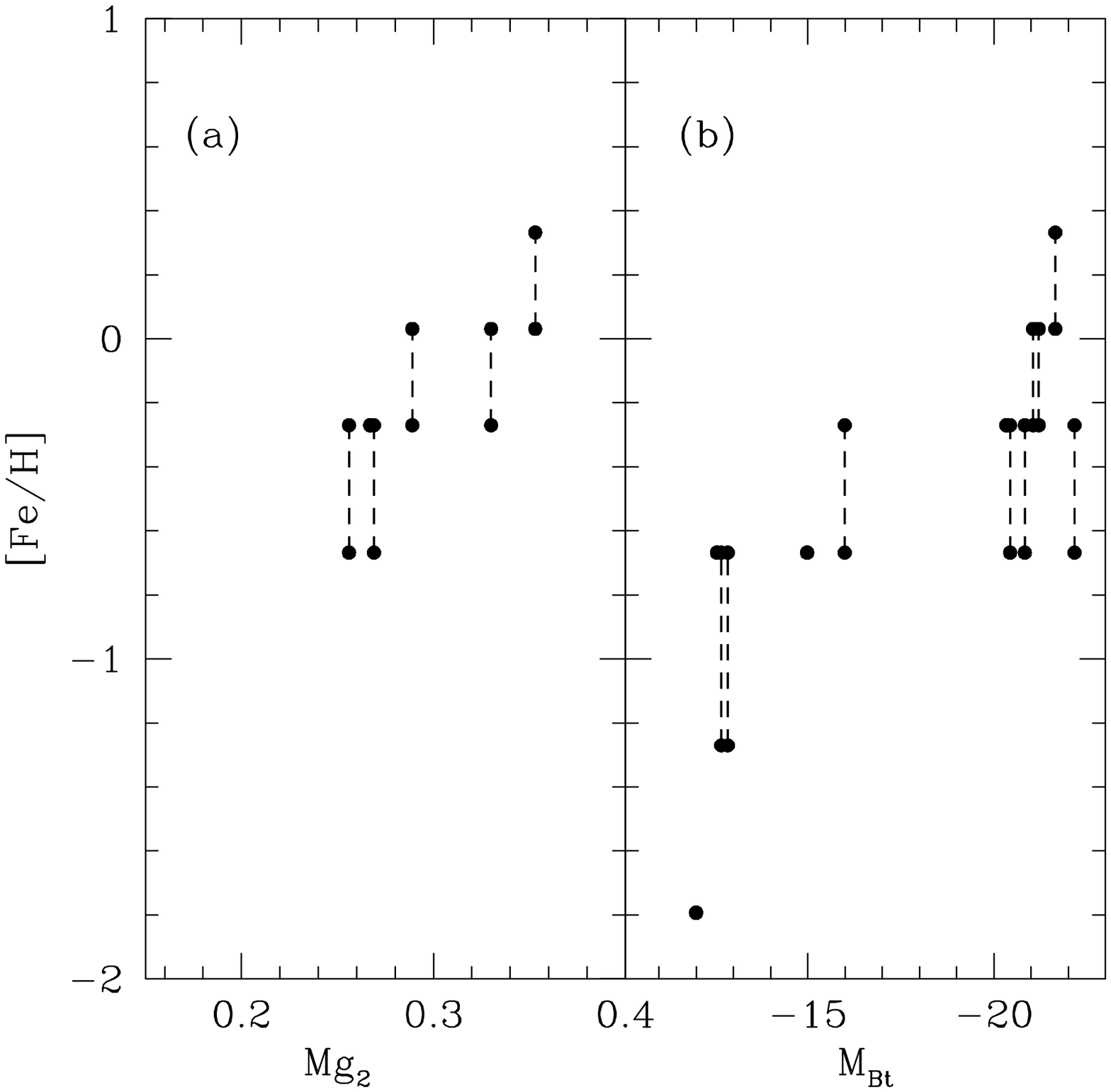}{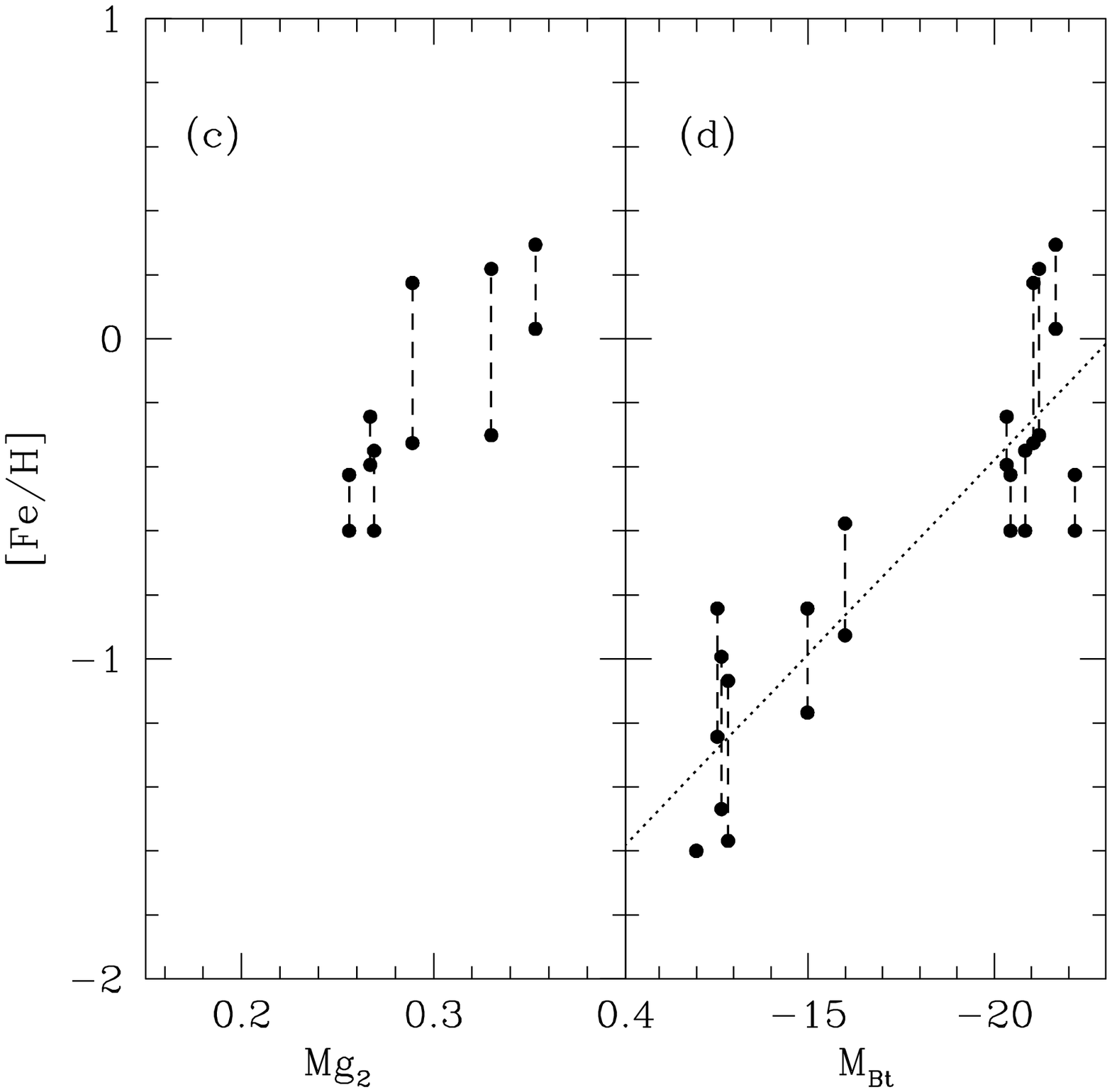}
\caption{Left panels: Minimum and maximum metallicity limits derived
using R05 populations synthesis models as a function of the $Mg_2$
(left), and of the total B magnitude $M_{Bt}$ of the galaxy
(right). Right panels: as left panels, but the average metallicity
derived from all models is considered (Table \ref{tab_ssp}). In panel
($d$) also a linear fit to the data is shown.
\label{mgmbfeh}}
\end{figure}

\begin{figure}[!ht]
\plottwo{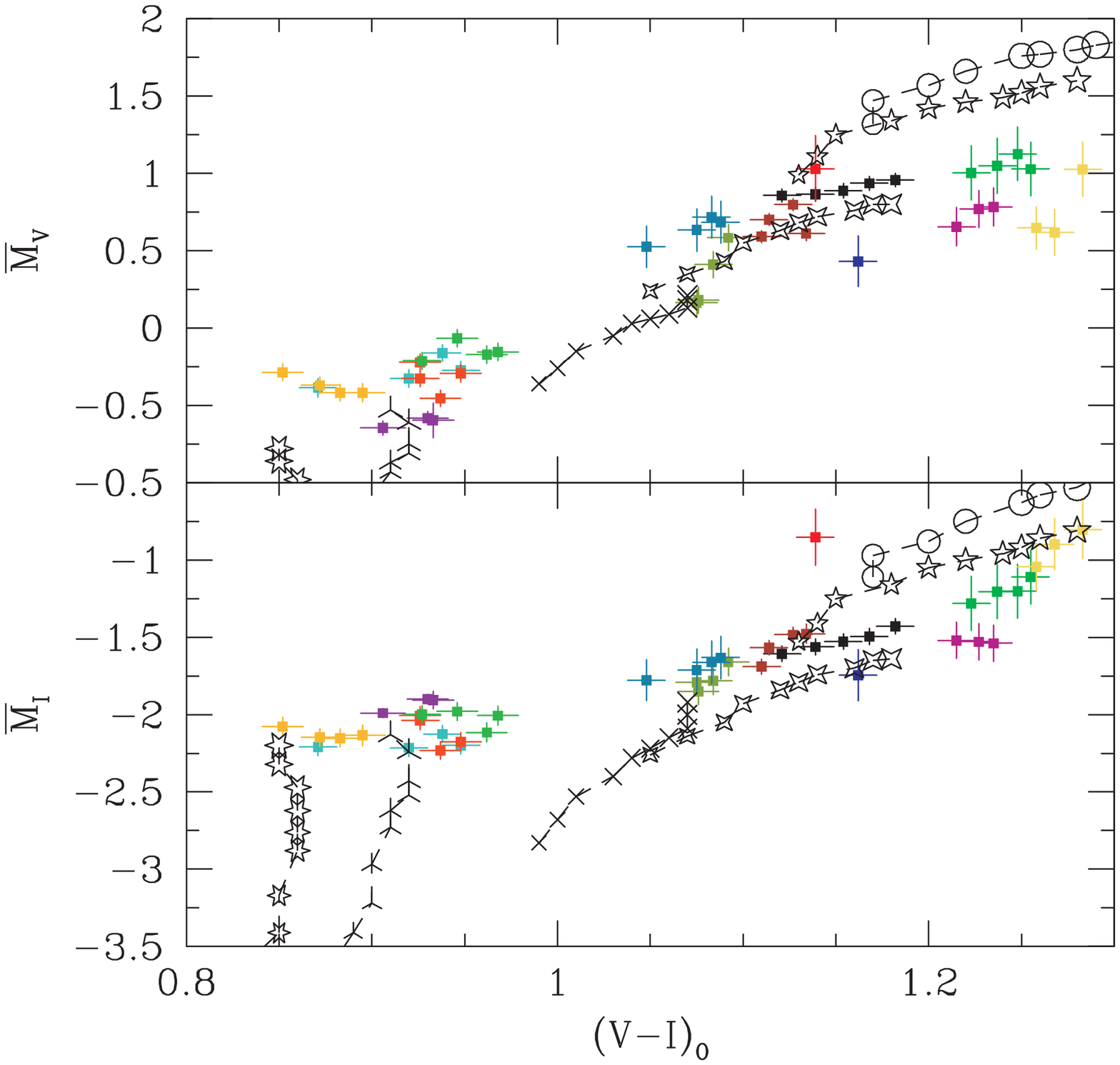}{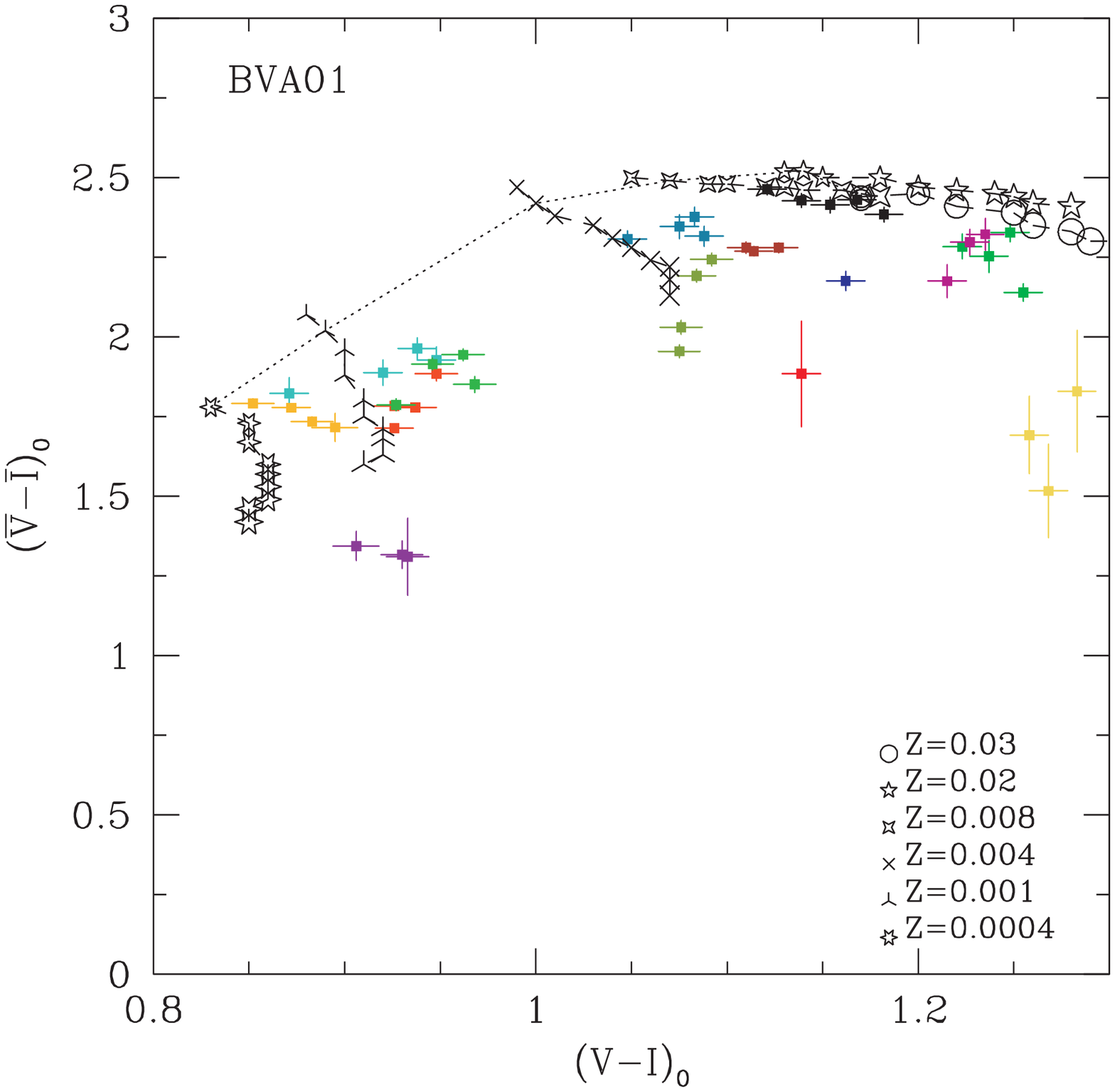}
\plottwo{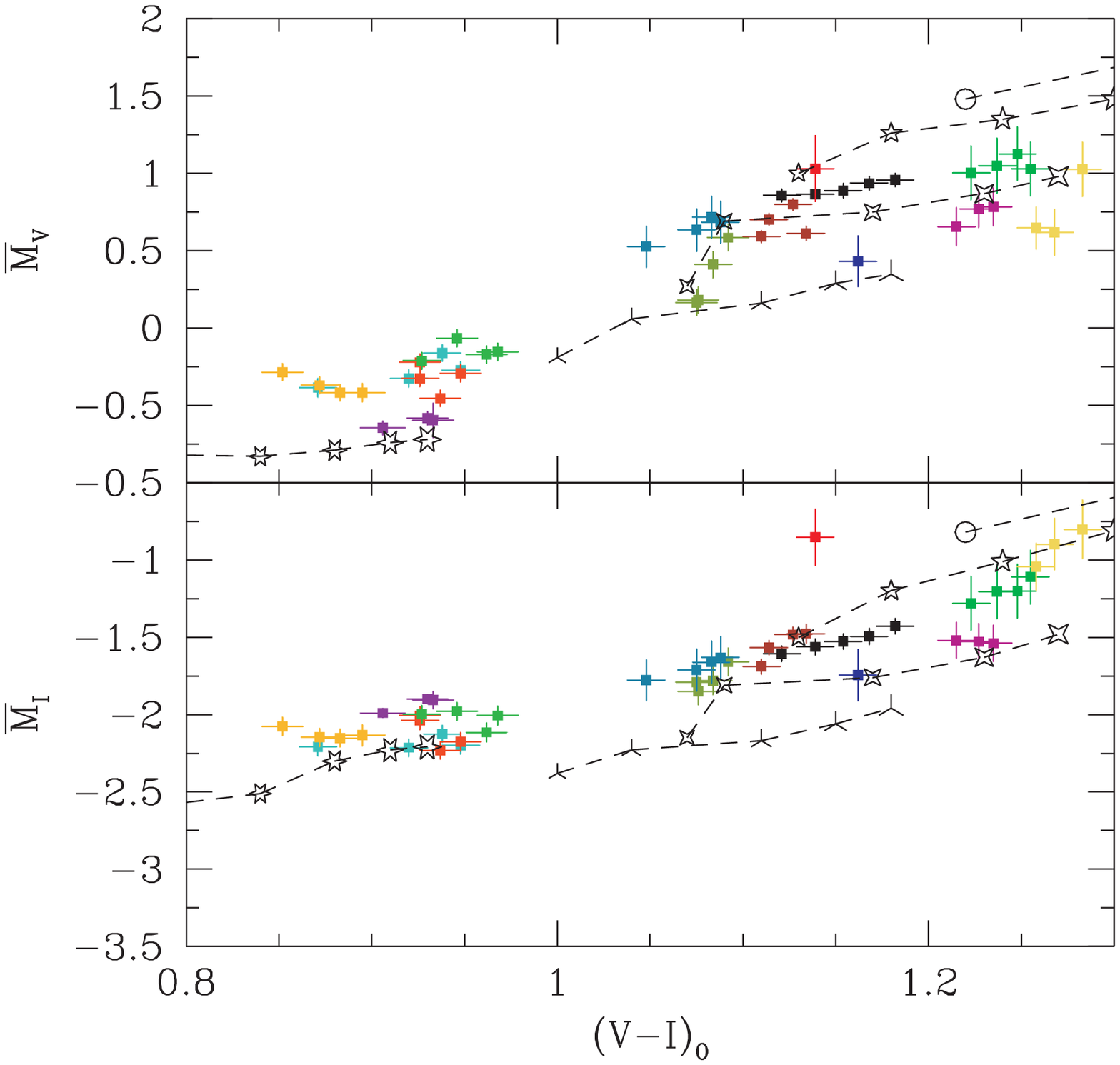}{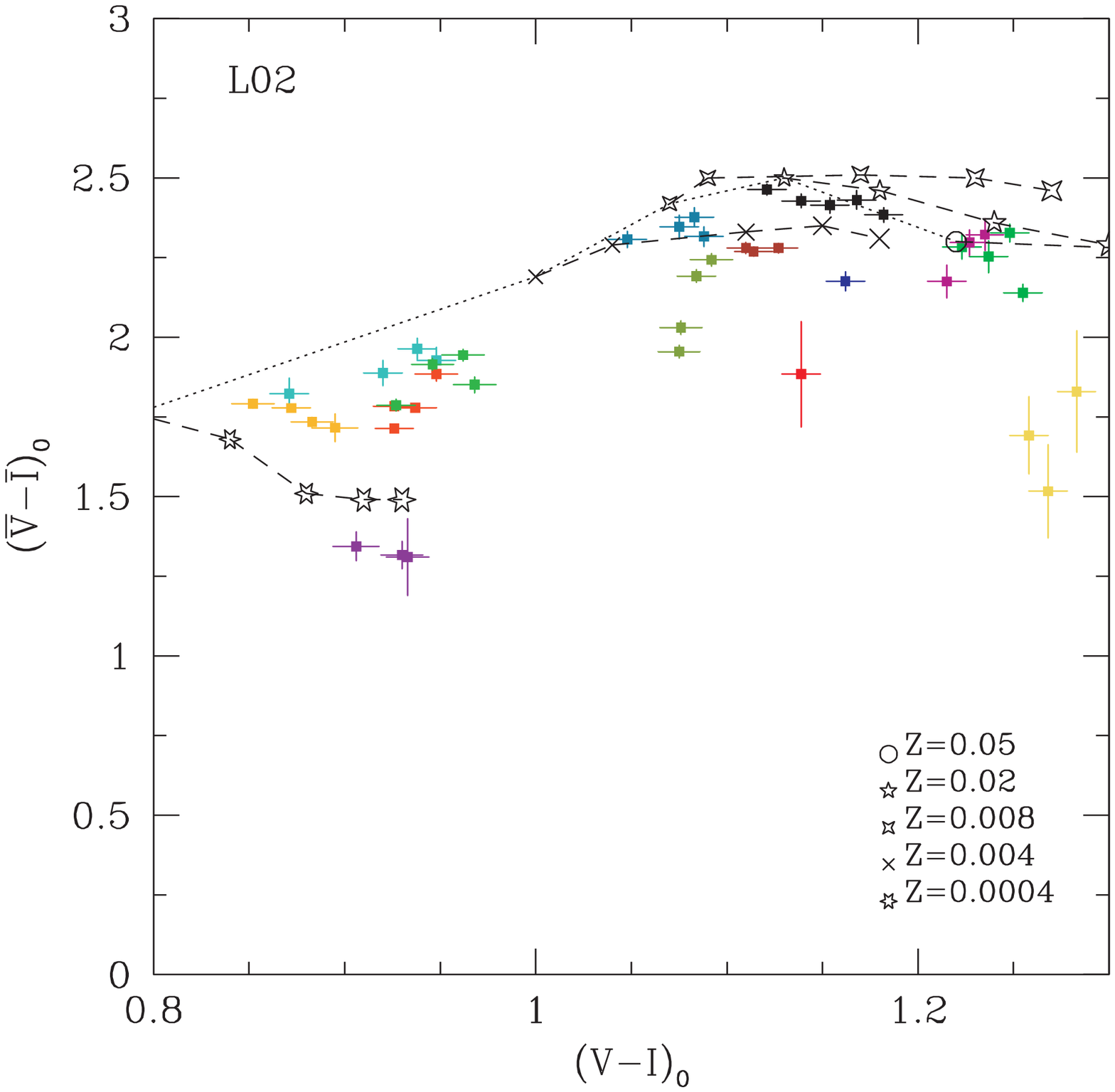}
\plottwo{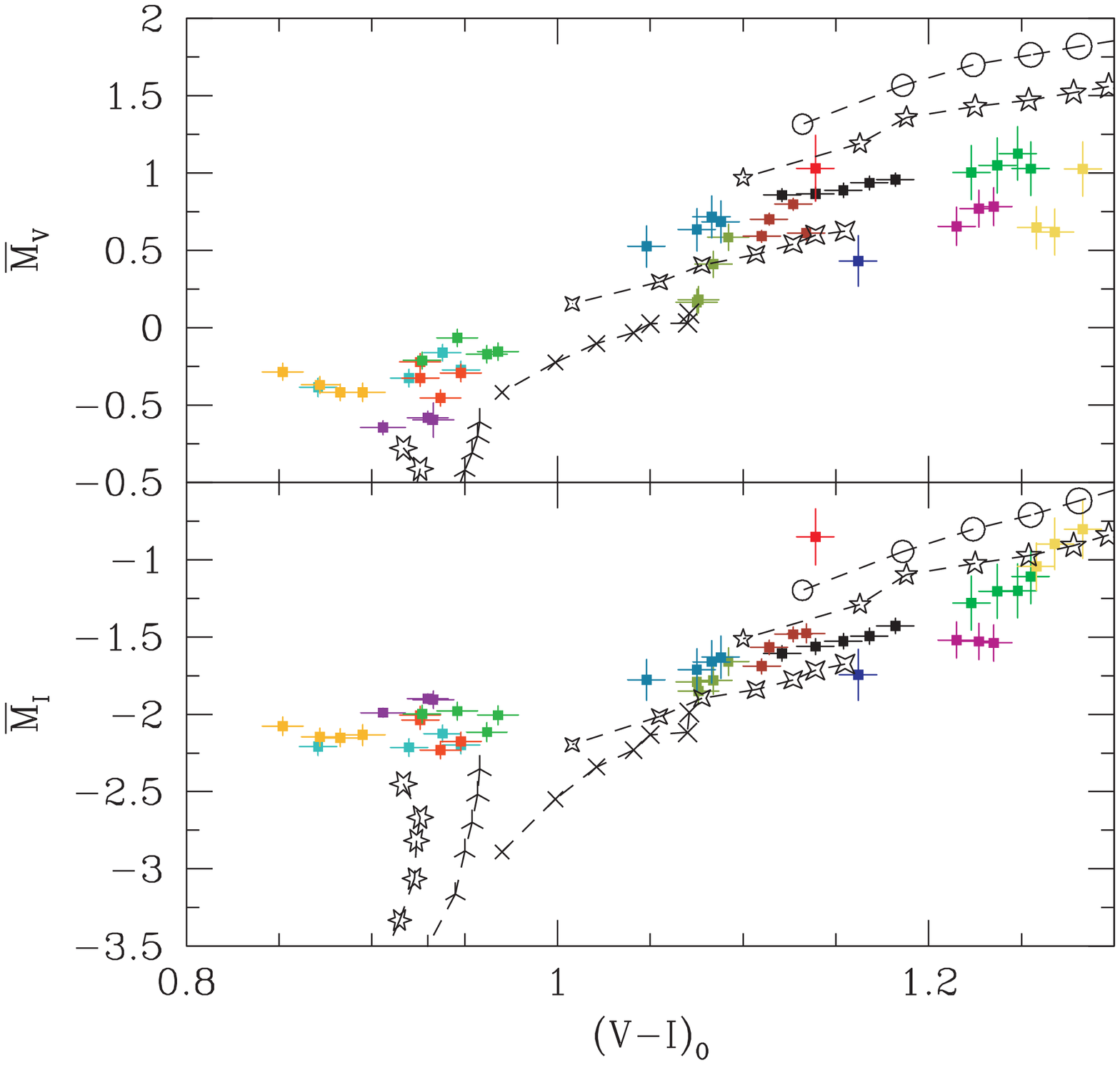}{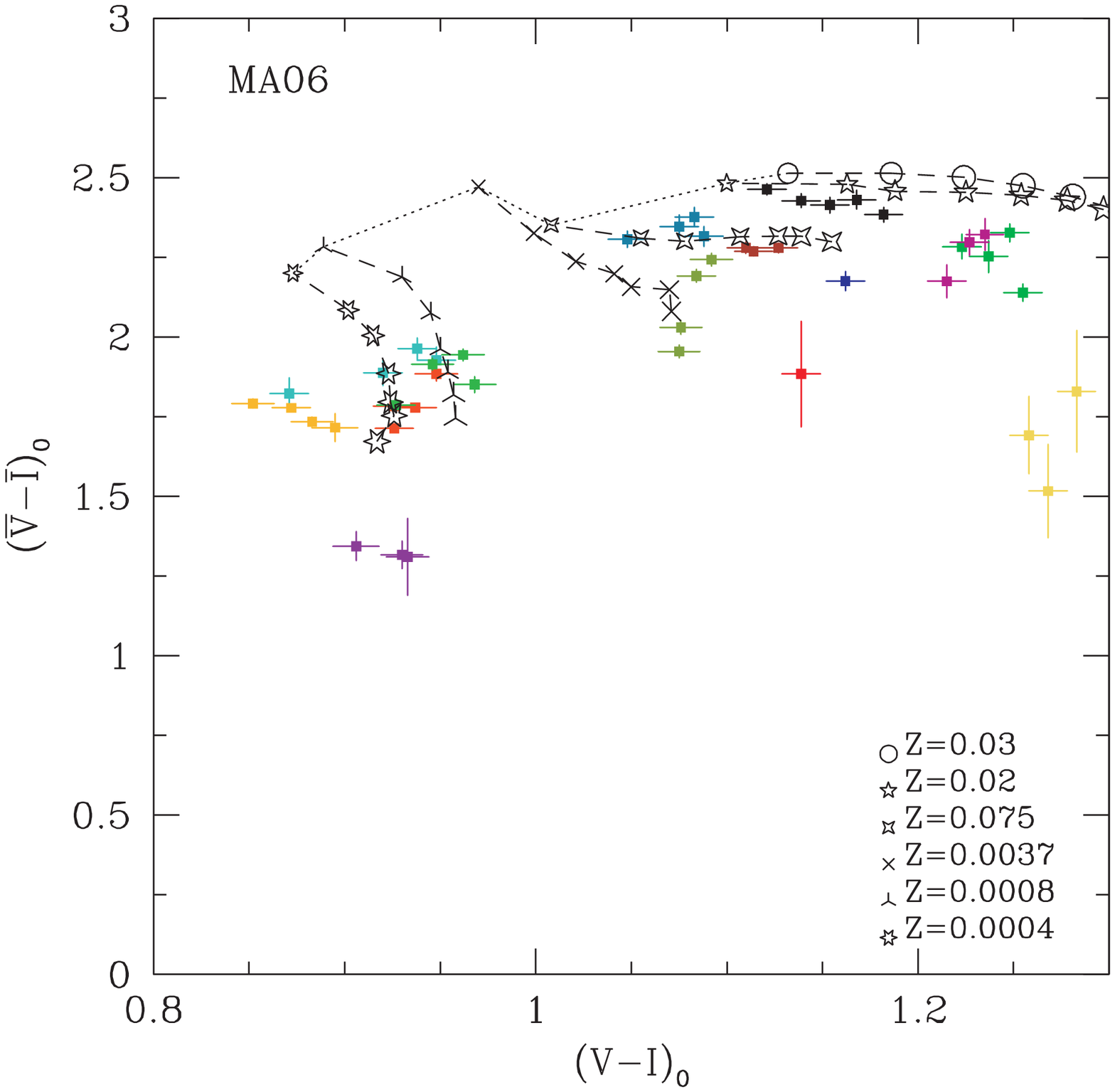}
\caption{As Figure \ref{spot} but for the BVA01 (upper panels), L02
(middle), and MA06 models (lower panels). The age range is t=5 to
t=17.8 Gyr (step of 12\%) for BVA01; t=3,5,8,12,17 Gyr for L02 models;
t=3,5,7,9,11,13,15 Gyr for MA06 models. The dotted lines connect
models at 5, 3, and 3 Gyr for the three models, respectively. 
\label{others}}
\end{figure}

\begin{itemize}
\item DDO\,71 -- This galaxy does not show an obvious gradient (Table
\ref{tab_measures}). On average, observational data are located in
between models of Z$\sim$ 0.004, and in the age interval 4-11 Gyr.
BVA01 and MA06 models predict older ages ($\gsim 10$ Gyr) respect to
L02 and R05.

\item KDG\,61 -- All models predict a $Z < 0.004$ stellar system.
Within the R05 and L02 SSP scenarios, the radial change of SBF and
integrated color could be interpreted by a radial change in the age of
the dominant stellar component. The opposite conclusion would be drawn
by using the BVA01 and MA06 models. Moreover, it must be noted the
criss-crossing of data at different radii for this galaxy.

\item KDG\,64 -- A 0.0004$\leq Z \leq 0.001$, t$>$ 5 Gyr system is
predicted. It is worth mentioning that for this galaxy, and for the
two previous -- all members of the M\,81 group -- \citet{dacosta07}
quotes an average chemical content of Z$\lsim$ 0.001 comparing the
mean RGB color to the colors of Galactic Globular Clusters. The
location of these galaxies in the low metallicity regime, where
significant discrepancies between models exist, does not allow us to
obtain any substantial conclusion on the possible origin of SBF versus
color gradients.
  
\item NGC\,474 -- The single measurement available for this galaxy
agrees with a Z$\sim$ 0.01, old (t$\gsim 14$ Gyr) stellar population,
for all models considered. Lower ages, and slightly higher metallicity
have been found using high S/N spectral analysis \citep{howell06},
though the spectral data refer to a smaller, more centrally
concentrated area compared to our measurements.

\item NGC\,1316 -- A Z$\sim 0.01$ is found from models, with age t$>$8
Gyr. The inner annuli seem to be populated by a rather old (t$\sim13$)
stellar system with respect to the outer annuli (t$\sim$8). Such
radial change of the stellar age would be also supported by the fact
that this galaxy is a known merger remnant \citep{goudfrooij01}.
  
\item NGC\,1344 -- Models to data comparison seems to point out a
 Z$\gsim 0.01$, t$>$5 Gyr stellar system. Also, for this galaxy all
 the models predict that the observed trend of SBF and color could be
 explained by an age gradient along the radius, with the inner regions
 being older than the outer ones. In fact, for this galaxy we find
 $\alpha \equiv \delta \bar{m}_I / \delta (V-I)_0 =2.8 \pm 0.3$, and all
 models predict $\alpha \equiv \delta \bar{M}_I / \delta (V-I)_0 \sim 3.5$ in
 this metallicity regime. As in the case of NGC\,1316, this galaxy
 shows indications of a recent merger activity \citep{carter82} which
 could possibly be related to the observed age gradient.

\item NGC\,2865 -- It is not unexpected that the only measurement
available for this galaxy is significantly out the grid of models,
no matter what set of SSP simulations is considered. In fact, as
shown in Figure \ref{single} and discussed in \S 4.1.2, this
galaxy has a peculiar behavior even when other physical properties are
taken into account.  With these caveats, NGC\,2865 data are located
between models of Z=0.004 and Z=0.01, on the side of the oldest
ages. Combining optical spectra and spectral synthesis
\citet{raimann05} have found that the light of this galaxy is mostly
dominated ($\sim$ 70 \% of the flux) by an t$\sim$10 Gyr stellar
system (these authors do not differentiate on metallicity). In spite
of such age agreement between the results obtained with two different
stellar population indicators, we must highlight that the V-band image
of NGC\,2865 clearly shows the presence of diffuse dust in this
galaxy. This finding, together with the peculiar behavior of this
galaxy with respect to other data from literature, lead us to reject
NGC\,2865 in the section dedicated to distance measurements.

\item NGC\,3610 -- Observational data match with models at 0.004 $\leq
Z\leq$0.01, in agreement with \citet{howell04}. The age range
predicted by different models is quite broad, going from $\sim$5 to
$\sim$ 16 Gyr. 

\item NGC\,3923 -- The metallicity predicted is Z$\sim$0.02, with old
ages (t$\gsim 10$ Gyr). The dominance of an old stellar population is
also found by \citet{raimann05}.

\item NGC\,5237 -- The chemical composition from models is generally
$Z\lsim 0.004$, with ages generally smaller than $\sim$11 Gyr.

\item NGC\,5982 -- A Z$\gsim$ 0.02, old stellar population is
invariably predicted for this galaxy. A comparable result is found by
\citet{denicolo05} from spectroscopic data -- though their data refer
to a smaller aperture.
 
\item NGC\,7626 -- This galaxy's data are significantly off the grid
of models.  A general feature that can be recognized from the location
of this galaxy in the SBF versus color panels, is that the stellar
population is very likely old, and metal rich, as also found by
\citet{denicolo05}. The galaxy shows the presence of a dust lane,
however we do not recognize irregular dust patches (neither from the
V-band image, nor from the B--band images also available from the ACS
archive) which might lead to mark as unreliable the SBF value for this
galaxy.

\item UGC\,7369 -- Data are consistent with a metallicity in the range
0.004 $\leq Z \leq $ 0.01, and ages typically t$\gsim 9$ Gyr. A
non--negligible preference on a metallicity gradient is
recognizable, with outer regions being more metal poor than the
inner ones.

\item VCC\,941 -- This galaxy's data match with models of very low
metallicity ($Z\sim 0.0004$), and old ages ($t\gsim$ 12 Gyr). The
SBF-gradient observed cannot be interpreted as related to age or
metallicity variations because of the substantial differences between
models in this metallicity regime. While BVA01 and MA06 models, in
fact, predict that the observed gradient might be related to
metallicity variations with the galaxy radius, the R05 and L02 are
much more consistent with age variations.
\end{itemize}

\clearpage

\clearpage


\begin{deluxetable}{cccccccccccccc}
\rotate
\tablecolumns{14}
\tabletypesize{\tiny}
\tablewidth{0pc}
\tablehead{
\colhead{Galaxy} & \colhead{R.A.} & \colhead{Decl.}  & \colhead{$v_{flow}$}  & \colhead{T}  & \colhead{Mg$_2$} & \colhead{$\sigma$ (Km s$^{-1}$)}  & \colhead{H$_{\beta}$}  &
\colhead{m$_{Bt}$} & \colhead{A$_B$}  & \colhead{$\mu_{0,group}$}  & \colhead{Ref.\tablenotemark{a}}     & \colhead{I}     & \colhead{V} }
\startdata
DDO\,71	  &  151.276 &   66.558 &    -102  & 10.0 $\pm$ 0.5 &    \nodata            &  \nodata       &     \nodata          &  15.69$\pm$ 0.25 &    0.412 &   27.84 $\pm$ 0.05 &   1 & 9000 & 17200\tablenotemark{b}  \\
KDG\,61	  &  149.262 &   68.591 &    -119  &  8.0 $\pm$ 3.0 &    \nodata            &  \nodata       &     \nodata         &  15.30$\pm$ 0.50 &    0.309 &   27.84 $\pm$ 0.05 &   1 & 9000 & 17200\tablenotemark{b}  \\
KDG\,64	  &  151.757 &   67.827 &     6   &  9.9 $\pm$ 0.6 &    \nodata             &  22.6$\pm$14.5 &    \nodata          &  15.40$\pm$ 0.50 &    0.235 &   27.84 $\pm$ 0.05 &   1 & 9000 & 17200\tablenotemark{b}  \\
NGC\,474 &   20.027 &    3.415 &    1899  & -2.0 $\pm$ 0.4 &     \nodata            &  163.9$\pm$5.1 &    1.71$\pm$  0.10   &  12.36$\pm$ 0.16 &    0.148 &   32.65 $\pm$ 0.16 &   2 &  960 &  1140\tablenotemark{b}  \\
NGC\,1316 &   50.673 &  -37.208 &   1441  & -1.7 $\pm$ 0.9 &     0.244$\pm$ 0.005   &  227.5$\pm$4.3 &    2.20$\pm$  0.07   &   9.40$\pm$ 0.28 &    0.090 &   31.48 $\pm$ 0.03 &   1 & 4850 &  6890\tablenotemark{c}  \\
NGC\,1344 &   52.081 &  -31.068 &    822  & -3.9 $\pm$ 1.4 &     0.257$\pm$ 0.003   &  166.4$\pm$4.1 &      \nodata         &  11.22$\pm$ 0.18 &    0.077 &   31.48 $\pm$ 0.03 &   1 &  960 &  1062\tablenotemark{b}  \\
NGC\,2865 &  140.875 &  -23.161 &   2737  & -4.1 $\pm$ 1.3 &     0.186$\pm$ 0.002   &  170.3$\pm$2.9 &    3.02$\pm$  0.16   &  12.45$\pm$ 0.22 &    0.355 &   32.91 $\pm$ 0.15 &   3 &  960 &  1020\tablenotemark{b}  \\
NGC\,3610 &  169.605 &   58.786 &   1819  & -4.2 $\pm$ 1.3 &     0.241$\pm$ 0.003   &  162.0$\pm$4.5 &    2.33$\pm$  0.06   &  11.61$\pm$ 0.13 &    0.043 &   32.47 $\pm$ 0.13 &   3 & 6060 &  6410\tablenotemark{c}  \\
NGC\,3923 &  177.757 &  -28.806 &   2060  & -4.6 $\pm$ 0.7 &     0.298$\pm$ 0.005   &  253.9$\pm$5.9 &    1.88$\pm$  0.07   &  10.78$\pm$ 0.14 &    0.357 &   31.72 $\pm$ 0.17 &   3 &  978 &  1140\tablenotemark{b}  \\
NGC\,5237 &  204.412 &  -42.846 &    710  &  1.4 $\pm$ 4.5 &    \nodata             &  \nodata       &     \nodata          &  13.23$\pm$ 0.07 &    0.414 &   27.80 $\pm$ 0.04 &   4 &  900 &  1200\tablenotemark{b}  \\
NGC\,5982 &  234.665 &   59.355 &   3198  & -4.8 $\pm$ 0.6 &     0.277$\pm$ 0.005   &  240.4$\pm$5.2 &    1.47$\pm$  0.07   &  11.98$\pm$ 0.12 &    0.077 &   33.38 $\pm$ 0.12 &   3 & 1020 &  1314\tablenotemark{b}  \\
NGC\,7626 &  350.176 &    8.217 &   3104  & -4.8 $\pm$ 0.5 &     0.337$\pm$ 0.002   &  274.0$\pm$4.9 &    1.33$\pm$  0.12   &  12.16$\pm$ 0.15 &    0.313 &   33.57 $\pm$ 0.11 &   3 &  960 &  1140\tablenotemark{b}  \\
UGC\,7369 &  184.911 &   29.883 &    522  &  6.0 $\pm$ 2.6 &  \nodata               &   \nodata      &     \nodata          &  15.16$\pm$ 0.41 &    0.083 &   31.07 $\pm$ 0.08 &   1 &  900 &  1200\tablenotemark{b}  \\
VCC\,941  &  186.698 &   13.380 &   1228  & -5.0 $\pm$ 3.0 &   \nodata              &   \nodata      &     \nodata          &  19.18$\pm$ 0.15 &    0.117 &   31.06 $\pm$ 0.03 &   1 & 14400& 33280\tablenotemark{c} \\
\enddata
\tablenotetext{a}{References -- (1) \citet{f00}, distance indicators adopted: Cepheids variables, RGB Tip, Planetary Nebulae Luminosity Function, 
Globular Clusters  Luminosity Function; (2) \citet{roberts91}: Hubble law ($H_0=$72 km $s^{-1}$ Mpc); (3) \citet{blake02}: Fundamental
Plane and IRAS redshift survey density field; (4) \citet{karachentsev02}: RGB Tip.}
\tablenotetext{b}{F606W passband data.}
\tablenotetext{c}{F555W passband data.}
\label{tab_dati}
\end{deluxetable}

\begin{deluxetable}{cccccccc}
\tablecolumns{8}
\tabletypesize{\tiny}
\tablewidth{0pc}
\tablecaption{SBF and color measurements}
\tablehead{\colhead{$\langle radius \rangle$} & \colhead{(V-I)$_0$} & \colhead{$P_{0,V}$} & \colhead{$P_{r,V}$} & \colhead{$\bar{V}_0$}  &  \colhead{$P_{0,I}$} & \colhead{$P_{r,I}$} & \colhead{$\bar{I}_0$}}
\startdata
\multicolumn{8}{c}{DDO\,71} \\
  9.3 &  0.968  0.020 &  6.14 $\cdot10^{3}$ & \nodata{\tablenotemark{a}} & 27.69   0.03 &  5.76 $\cdot10^{3}$ & \nodata{\tablenotemark{a}} & 25.83   0.03  \\ 
 15.1 &  0.962  0.020 & 6.23  $\cdot10^{3}$ & \nodata & 27.67   0.02 &  6.39  $\cdot10^{3}$ & \nodata & 25.73   0.03  \\ 
 20.4 &  0.946  0.020 & 5.72  $\cdot10^{3}$ & \nodata & 27.77   0.02 &  5.64  $\cdot10^{3}$ &  \nodata &25.86   0.03  \\ 
 28.6 &  0.927  0.019 & 6.40  $\cdot10^{3}$ & \nodata & 27.63   0.02 &  5.71  $\cdot10^{3}$ & \nodata & 25.84   0.03  \\ 
$\langle av. \rangle_w$ & 0.950 0.010& \nodata & \nodata & 27.69  0.01 & \nodata & \nodata &  25.82   0.02  \\  
\multicolumn{8}{c}{$\alpha= -1.5$, $\Delta \alpha =2.1$ : SBF-flat} \\
\hline
\multicolumn{8}{c}{KDG\,61} \\
  4.2 &  0.895  0.019 &  8.09 $\cdot10^{3}$ & \nodata{\tablenotemark{a}} & 27.42   0.03 & 6.70  $\cdot10^{3}$ & \nodata{\tablenotemark{a}} & 25.71   0.04  \\ 
  8.4 &  0.852  0.018 &  7.26 $\cdot10^{3}$ & \nodata & 27.55   0.02 & 6.37  $\cdot10^{3}$ & \nodata & 25.76   0.03  \\ 
 14.6 &  0.883  0.019 &  8.12 $\cdot10^{3}$ & \nodata & 27.42   0.02 & 6.81  $\cdot10^{3}$ & \nodata & 25.69   0.03  \\ 
 19.1 &  0.872  0.019 &  7.83 $\cdot10^{3}$ & \nodata & 27.47   0.02 & 6.80  $\cdot10^{3}$ & \nodata & 25.69   0.03  \\ 
$\langle av. \rangle_w$ &  0.875  0.009 &  \nodata & \nodata & 27.47   0.01 &  \nodata & \nodata & 25.71   0.02  \\  
\multicolumn{8}{c}{$\alpha= -1.4$, $\Delta \alpha =0.9$: SBF-flat} \\
\hline
\multicolumn{8}{c}{KDG\,64} \\
  9.6 &  0.948  0.020 & 7.78  $\cdot10^{3}$ &  \nodata{\tablenotemark{a}} & 27.55   0.03 & 7.22 $\cdot10^{3}$ & \nodata{\tablenotemark{a}} & 25.66   0.03  \\ 
 12.3 &  0.926  0.019 & 7.17  $\cdot10^{3}$ &  \nodata & 27.62   0.02 & 6.15 $\cdot10^{3}$ & \nodata & 25.83   0.03  \\ 
 17.3 &  0.937  0.019 & 8.85  $\cdot10^{3}$ &  \nodata & 27.39   0.02 & 7.58 $\cdot10^{3}$ & \nodata & 25.61   0.03  \\ 
 25.3 &  0.926  0.019 & 7.77  $\cdot10^{3}$ &  \nodata & 27.52   0.02 & 6.33 $\cdot10^{3}$ & \nodata & 25.80   0.03  \\ 
$\langle av. \rangle_w$ &  0.934  0.010 & \nodata & \nodata & 27.51   0.01 & \nodata & \nodata & 25.74   0.01  \\  
\multicolumn{8}{c}{$\alpha= -8.1$, $\Delta \alpha =4.5$: SBF-flat} \\
\hline
\multicolumn{8}{c}{NGC\,474} \\
 18.2 &  1.162  0.022 &  3.62 & 0.13 & 33.08   0.03 &  6.60 & 0.13 & 30.91   0.04  \\ 
\hline
\multicolumn{8}{c}{NGC\,1316} \\
 36.3 &  1.134  0.021 &  16.24 & 0.08 & 32.09   0.03 &   \nodata & \nodata & \nodata       \\ 
 42.1 &  1.127  0.021 &  13.60 & 0.09 & 32.28   0.02 &   77.67   &  0.12   & 30.00   0.04  \\ 
 58.6 &  1.114  0.021 &  14.90 & 0.14 & 32.18   0.02 &   83.85   &  0.15   & 29.91   0.04  \\ 
 74.9 &  1.110  0.020 &  16.58 & 0.23 & 32.07   0.02 &   94.07   &  0.23   & 29.79   0.04  \\ 
$\langle av. \rangle_w$ &  1.121  0.010 & \nodata & \nodata & 32.19   0.01 &  \nodata & \nodata & 29.91   0.02  \\  
\multicolumn{8}{c}{$\alpha=8.0$, $\Delta \alpha =2.7$: SBF-gradient} \\
\hline
\multicolumn{8}{c}{NGC\,1344} \\
 13.8 &  1.182  0.023 &  6.41 & 0.08 & 32.44   0.03 &  14.78 & 0.07 & 30.05   0.04  \\ 
 19.1 &  1.168  0.022 &  6.58 & 0.08 & 32.42   0.03 &  15.75 & 0.07 & 29.98   0.04  \\ 
 27.0 &  1.154  0.022 &  6.87 & 0.10 & 32.37   0.03 &  16.22 & 0.08 & 29.95   0.04  \\ 
 37.3 &  1.139  0.022 &  7.07 & 0.14 & 32.35   0.03 &  16.81 & 0.12 & 29.92   0.04  \\ 
 44.0 &  1.121  0.022 &  7.27 & 0.25 & 32.34   0.03 &  17.63 & 0.21 & 29.87   0.04  \\ 
$\langle av. \rangle_w$ &  1.152  0.010 & \nodata & \nodata & 32.38   0.01 &  \nodata & \nodata & 29.96   0.02  \\  
\multicolumn{8}{c}{$\alpha= 2.8$, $\Delta \alpha =0.3$: SBF-gradient} \\
\hline
\multicolumn{8}{c}{NGC\,2865} \\
 17.6 &  1.139  0.022 &  1.39 & 0.19 & 33.94   0.15 &  2.24 & 0.20 & 32.06   0.10  \\ 
\hline
\multicolumn{8}{c}{NGC\,3610} \\
 13.3 &  1.088  0.020 &  5.84 & 0.14 & 33.16   0.03 &  46.02 & 0.59 & 30.84   0.04  \\ 
 18.5 &  1.083  0.020 &  5.67 & 0.14 & 33.19   0.03 &  47.20 & 0.38 & 30.81   0.04  \\ 
 28.0 &  1.075  0.020 &  6.19 & 0.23 & 33.10   0.04 &  49.38 & 0.38 & 30.76   0.04  \\ 
 47.0 &  1.048  0.014 &  7.20 & 0.61 & 32.99   0.03 &  52.77 & 0.73 & 30.69   0.03  \\ 
$\langle av. \rangle_w$ &  1.068  0.009 &  \nodata & \nodata & 33.10   0.02 & \nodata & \nodata & 30.76   0.02  \\  
\multicolumn{8}{c}{$\alpha= 3.5$, $\Delta \alpha =0.6$: SBF-gradient} \\
\hline
\multicolumn{8}{c}{NGC\,3923} \\
 13.5 &  1.255  0.024 &  4.26 & 0.09 & 32.75   0.03 & 8.07 & 0.11 & 30.61   0.04  \\ 
 18.4 &  1.248  0.024 &  4.02 & 0.09 & 32.85   0.03 & 8.82 & 0.10 & 30.52   0.04  \\ 
 26.7 &  1.237  0.023 &  4.26 & 0.10 & 32.77   0.05 & 8.81 & 0.10 & 30.52   0.04  \\ 
 37.4 &  1.223  0.023 &  4.49 & 0.12 & 32.72   0.04 & 9.48 & 0.11 & 30.44   0.04  \\ 
$\langle av. \rangle_w$ &  1.240  0.012 &  \nodata & \nodata & 32.78   0.02 &  \nodata & \nodata & 30.53   0.02  \\  
\multicolumn{8}{c}{$\alpha= 4.6$, $\Delta \alpha =1.3$: SBF-gradient} \\
\hline
\multicolumn{8}{c}{NGC\,5237} \\
  7.6 &  0.871  0.019 & 546.7 & 0.8 & 27.41   0.05 &  721.4 & 2.4 & 25.59   0.04  \\ 
 13.4 &  0.920  0.019 & 523.7 & 0.6 & 27.47   0.04 &  725.3 & 1.3 & 25.59   0.04  \\ 
 18.9 &  0.938  0.019 & 456.1 & 0.6 & 27.64   0.04 &  670.6 & 1.3 & 25.68   0.04  \\ 
 27.1 &  0.948  0.020 & 502.7 & 0.8 & 27.53   0.04 &  718.0 & 1.7 & 25.60   0.04  \\ 
$\langle av. \rangle_w$ &  0.919  0.010 &  \nodata & \nodata & 27.53   0.02 &   \nodata & \nodata & 25.62   0.02  \\  
\multicolumn{8}{c}{$\alpha= 0.5$, $\Delta \alpha =0.8$: SBF-flat} \\
\hline
\multicolumn{8}{c}{NGC\,5982} \\
 13.7 &  1.235  0.023 &  1.71 & 0.12 & 34.16   0.06 & 3.12 & 0.13 & 31.84   0.04  \\ 
 19.2 &  1.227  0.023 &  1.74 & 0.13 & 34.15   0.04 & 3.10 & 0.13 & 31.85   0.04  \\ 
 27.8 &  1.215  0.023 &  1.93 & 0.18 & 34.03   0.06 & 3.11 & 0.18 & 31.86   0.04  \\ 
$\langle av. \rangle_w$ &  1.226  0.013 &  \nodata & \nodata &  34.12   0.03 &  \nodata & \nodata &  31.85   0.02  \\  
\multicolumn{8}{c}{$\alpha= -0.9$, $\Delta \alpha =0.1$: SBF-gradient} \\
\hline
\multicolumn{8}{c}{NGC\,7626} \\
  9.5 &  1.283  0.024 &  1.05 & 0.31 & 34.60   0.14 & 1.58 & 0.50 & 32.77   0.15  \\ 
 15.2 &  1.268  0.024 &  1.25 & 0.24 & 34.19   0.10 & 1.49 & 0.33 & 32.67   0.12  \\ 
 21.1 &  1.258  0.024 &  1.26 & 0.25 & 34.22   0.08 & 1.62 & 0.29 & 32.52   0.10  \\ 
$\langle av. \rangle_w$ &  1.270  0.014 & \nodata & \nodata &   34.27   0.06 & \nodata & \nodata &  32.62   0.07  \\  
\multicolumn{8}{c}{$\alpha= 9.4$, $\Delta \alpha =2.3$: SBF-gradient} \\
\hline
\multicolumn{8}{c}{UGC\,7369} \\
  4.3 &  1.092  0.021 & 14.5 & 0.1 & 31.65   0.03 & 25.0 & 0.3 &  29.41   0.04  \\ 
  8.9 &  1.084  0.021 & 16.9 & 0.1 & 31.48   0.03 & 27.8 & 0.2 &  29.29   0.04  \\ 
 14.9 &  1.076  0.021 & 20.4 & 0.3 & 31.25   0.03 & 29.4 & 0.3 &  29.22   0.04  \\ 
 20.4 &  1.075  0.021 & 20.7 & 0.5 & 31.23   0.02 & 27.9 & 0.5 &  29.28   0.03  \\ 
$\langle av. \rangle_w$ &  1.082  0.011 &  \nodata & \nodata &  31.39   0.01 &  \nodata & \nodata &  29.30   0.02  \\  
\multicolumn{8}{c}{$\alpha= 9.0$, $\Delta \alpha =3.2$: SBF-gradient} \\
\hline
\multicolumn{8}{c}{VCC\,941} \\
  2.0 &  0.933  0.018 &  3.5 $\cdot10^{2}$ & \nodata{\tablenotemark{a}} & 30.48   0.11 & 4.7 $\cdot10^{2}$ & \nodata{\tablenotemark{a}} &  29.17   0.05  \\ 
  4.1 &  0.930  0.018 &  3.4 $\cdot10^{2}$ & \nodata & 30.50   0.04 & 4.7 $\cdot10^{2}$ & \nodata &  29.18   0.03  \\ 
  8.2 &  0.906  0.018 &  3.6 $\cdot10^{2}$ & \nodata & 30.43   0.04 & 5.1 $\cdot10^{2}$ & \nodata &  29.09   0.03  \\ 
$\langle av. \rangle_w$ &  0.923  0.010 &  \nodata & \nodata & 30.47   0.03 &  \nodata & \nodata & 29.14   0.02  \\  
\multicolumn{8}{c}{$\alpha= 3.4$, $\Delta \alpha =0.6$: SBF-gradient} \\
\enddata
\tablenotetext{a}{For these galaxies no correction for variance from external sources has been applied (see text). The few GC  present in these dwarf
galaxies have been masked out. Concerning background galaxies, after masking the brighter sources, the contribution
to the fluctuations of the fainter objects is by all means small respect to the stellar fluctuations. In fact the $P_r / (P_0 - P_r) \equiv P_r /P_f$ ratio 
due to the faint undetected galaxies is $<$ 0.001, that is the $P_r$ correction has no practical
effects on the final measured SBF.}
\label{tab_measures}
\end{deluxetable}

\begin{deluxetable}{cccccccc}
\tablecolumns{8}
\tabletypesize{\scriptsize}
\tablewidth{0pc}
\tablecaption{Stellar populations properties obtained using the different sets of models}
\tablehead{
\colhead{Galaxy} &  \colhead{R05} & \colhead{BVA01} & \colhead{L02}  & \colhead{MA06} \\
\colhead{(1)} &  \colhead{(2)} & \colhead{(3)} & \colhead{(4)}  & \colhead{(5)}}
\startdata
DDO\,71    &   Z$\sim$ 0.004          &  0.001 $\leq Z \leq$ 0.004  &  0.0004 $\leq Z \leq$0.004   &     Z $\sim$0.0008             \\
\nodata    &    4 $\leq t \leq$ 11    &   10 $\leq t \leq$ 12.6     &    5 $\leq t \leq$ 8         &     9 $\leq t \leq$ 13         \\         
\hline	                             
KDG\,61    &0.001 $\leq Z\leq$0.004   &  0.0004 $\leq Z \leq$ 0.001 &  0.001 $\leq Z \leq$0.004    &    Z $\sim$0.0004              \\
\nodata    &   3 $\leq t \leq$ 9      &   5 $\leq t \leq$ 12.6      &    3 $\leq t \leq$ 8         &    5 $\leq t \leq$ 15          \\     
\hline	                             
KDG\,64    &0.001  $\leq Z\leq$0.004  &  Z $\sim$ 0.001             &  0.0004 $\leq Z \leq$0.004   &    0.0004 $\leq Z \leq$0.0008  \\
\nodata    &   5 $\leq t \leq$ 11     &   11.2 $\leq t \leq$ 14.1   &    5 $\leq t \leq$ 8         &    9 $\leq t \leq$ 15          \\     
\hline	                             
NGC\,474  &0.004  $\leq Z\leq$0.01   &  0.004 $\leq Z \leq$ 0.008  &             Z $\sim$0.004    &    Z $\sim$ 0.0075              \\
\nodata    &   t $\sim$ 14            &   t $>$ 17.8                &           t $\sim$ 17        &    t $\sim$ 15                  \\     
\hline	                             
NGC\,1316  &0.004  $\leq Z\leq$0.01   &  0.004 $\leq Z \leq$ 0.008  &             Z $\sim$0.004    &    Z $\sim$ 0.0075             \\
\nodata    &  t $\sim$ 14             &   t $>$ 12.6                &      8 $\leq t \leq$ 12      &    9 $\leq t \leq$ 13          \\     
\hline	                             
NGC\,1344  &            Z$\sim$0.01   &  Z $\gsim$ 0.008            &    0.004 $\leq Z \leq$0.008  &    Z $\sim$ 0.02               \\
\nodata    &  9 $\leq t \leq$ 14      &  10 $\leq t \leq$ 17.8      &  8 $\leq t \leq$ 17          &    3 $\leq t \leq$ 11          \\     
\hline	                             
NGC\,2865  & 0.004  $\leq Z\leq$0.01  &   0.004 $\leq Z \leq$ 0.008 &    0.0004 $\leq Z \leq$0.004 &   0.0037 $\leq Z \leq$0.0075   \\
\nodata    &    t $\gsim$ 14          &    t $\gsim$ 17.8           &     t $\gsim$ 17             &   t $\gsim$ 15                 \\
\hline	                             
NGC\,3610  &0.004  $\leq Z\leq$0.01   &  0.004 $\leq Z \leq$ 0.008  &   0.004 $\leq Z \leq$0.008   &    Z $\gsim$ 0.0075           \\
\nodata    &  5  $\leq t \leq$ 13     &    11.2 $\leq t \leq$ 15.8  &      5 $\leq t \leq$ 8       &    5 $\leq t \leq$ 9          \\     
\hline	                             
NGC\,3923  & 0.01 $\leq Z \leq$0.02   &  0.01 $\leq Z \leq$ 0.03    &   0.01 $\leq Z \leq$0.05     &   0.0075 $\leq Z \leq$0.03     \\
\nodata    &  t $\sim$ 14             &    14.1 $\leq t \leq$ 17.8  &      3 $\leq t \leq$ 17      &     t $\sim$ 15                \\     
\hline	                             
NGC\,5237  &  Z$\lsim$0.004           &  Z $\sim$ 0.004             &   0.0004 $\leq Z \leq$0.004  &    Z $\lsim$ 0.0008           \\
\nodata    &  3  $\leq t \leq$ 7      &    7.9 $\leq t \leq$ 11.2   &      5 $\leq t \leq$ 8       &    7 $\leq t \leq$ 11          \\     
\hline	                             
NGC\,5982  & 0.01 $\leq Z \leq$ 0.02  &  0.01 $\leq Z \leq$ 0.02    &   0.008 $\leq Z \leq$0.05    &    0.0075 $\leq Z \leq$0.03    \\
\nodata    &  t $\geq$ 14             &    t $\geq$ 14.1            &      3 $\leq t \leq$ 17      &    t $\geq$ 15                 \\     
\hline	                             
NGC\,7626  & 0.02 $\leq Z \leq$ 0.04  &   0.02 $\leq Z \leq$ 0.03   &   0.02 $\leq Z \leq$ 0.05    &   0.02 $\leq Z \leq$ 0.03      \\
\nodata    &    t $\gsim$ 14          &    t $\gsim$ 17.8           &     t $\gsim$ 17             &   t $\gsim$ 15                 \\
\hline	                             
UGC\,7639  & 0.004  $\leq Z\leq$0.01  &  Z $\sim$ 0.004             &   0.0004 $\leq Z \leq$0.004  &    Z $\sim$ 0.0037              \\
\nodata    &  t $\sim$ 14             &    12.6 $\leq t \gsim$ 17.8 &      8 $\leq t \leq$ 17      &  9 $\leq t \leq$ 15             \\     
\hline	                             
VCC\,941   & Z$\sim$0.0003            &  Z $\lsim$ 0.001            &   Z $\sim$0.0004             &   Z $<$ 0.0008                 \\
\nodata    &  13  $\leq t \leq$ 14    &   t $\gsim$ 17.8            &      12 $\leq t \leq$ 17     &   t $\gsim$ 15                 \\     
\enddata
\label{tab_ssp}
\end{deluxetable}

\begin{deluxetable}{cccccccc}
\tablecolumns{8}
\tabletypesize{\scriptsize}
\rotate
\tablewidth{0pc}
\tablecaption{Distance Moduli with the different calibrations}
\tablehead{
\colhead{Galaxy} & \colhead{(V-I)$_0$} & \colhead{$\mu_{0,group}$}  & \colhead{$\mu_{0,T01}$}  & \colhead{$\mu_{0,J03}$}  & 
\colhead{$\mu_{0,M06}$} & \colhead{$\mu_{0,BVA01}$}  & \colhead{$\mu_{0,Ave}$\tablenotemark{a}} \\
\colhead{(1)} & \colhead{(2)} & \colhead{(3)}  & \colhead{(4)}  & \colhead{(5)}  & 
\colhead{(6)} & \colhead{(7)}  & \colhead{(8)}} 
\startdata
DDO\,71   &  0.950 $\pm$  0.010 & 27.84 $\pm$  0.05 & 27.98 $\pm$  0.27 & 27.98 $\pm$  0.27 & 28.31 $\pm$  0.20 & 28.18 $\pm$  0.27 & 28.08 $\pm$  0.19 \\
KDG\,61   &  0.875 $\pm$  0.009 & 27.84 $\pm$  0.05 & 27.88 $\pm$  0.27 & 27.88 $\pm$  0.27 & 28.39 $\pm$  0.30 & 27.97 $\pm$  0.27 & 27.92 $\pm$  0.19 \\
KDG\,64   &  0.934 $\pm$  0.010 & 27.84 $\pm$  0.05 & 27.90 $\pm$  0.27 & 27.90 $\pm$  0.27 & 28.27 $\pm$  0.21 & 28.00 $\pm$  0.27 & 27.95 $\pm$  0.19 \\
NGC\,474  &  1.162 $\pm$  0.022 & 32.65 $\pm$  0.16 & 32.59 $\pm$  0.13 & 32.43 $\pm$  0.13 & 32.43 $\pm$  0.13 & 32.21 $\pm$  0.17 & 32.35 $\pm$  0.11 \\
NGC\,1316 &  1.121 $\pm$  0.010 & 31.48 $\pm$  0.03 & 31.78 $\pm$  0.09 & 31.62 $\pm$  0.09 & 31.62 $\pm$  0.09 & 31.53 $\pm$  0.13 & 31.59 $\pm$  0.08 \\
NGC\,1344 &  1.152 $\pm$  0.010 & 31.48 $\pm$  0.03 & 31.69 $\pm$  0.09 & 31.53 $\pm$  0.09 & 31.53 $\pm$  0.09 & 31.56 $\pm$  0.13 & 31.54 $\pm$  0.08 \\
NGC\,3610 &  1.068 $\pm$  0.009 & 32.47 $\pm$  0.13 & 32.47 $\pm$  0.13 & 32.71 $\pm$  0.09 & 32.97 $\pm$  0.22 & 32.72 $\pm$  0.15 & 32.71 $\pm$  0.08 \\
NGC\,3923 &  1.240 $\pm$  0.012 & 31.72 $\pm$  0.17 & 31.86 $\pm$  0.10 & 31.70 $\pm$  0.10 & 31.70 $\pm$  0.10 & 31.49 $\pm$  0.16 & 31.64 $\pm$  0.08 \\
NGC\,5237 &  0.919 $\pm$  0.010 & 27.80 $\pm$  0.04 & 27.78 $\pm$  0.27 & 27.78 $\pm$  0.27 & 28.19 $\pm$  0.23 & 28.02 $\pm$  0.27 & 27.90 $\pm$  0.19 \\
NGC\,5982 &  1.226 $\pm$  0.013 & 33.38 $\pm$  0.12 & 33.25 $\pm$  0.10 & 33.09 $\pm$  0.10 & 33.09 $\pm$  0.10 & 32.91 $\pm$  0.15 & 33.03 $\pm$  0.09 \\
NGC\,7626 &  1.270 $\pm$  0.014 & 33.57 $\pm$  0.11 & 33.83 $\pm$  0.13 & 33.67 $\pm$  0.13 & 33.67 $\pm$  0.13 & 32.83 $\pm$  0.18 & 33.38 $\pm$  0.10 \\
UGC\,7369 &  1.082 $\pm$  0.011 & 31.07 $\pm$  0.08 & 31.35 $\pm$  0.10 & 31.19 $\pm$  0.10 & 31.47 $\pm$  0.23 & 30.94 $\pm$  0.14 & 31.11 $\pm$  0.08 \\
VCC\,941  &  0.923 $\pm$  0.010 & 31.06 $\pm$  0.03 & 31.31 $\pm$  0.27 & 31.31 $\pm$  0.27 & 31.70 $\pm$  0.23 & 30.96 $\pm$  0.27 & 31.14 $\pm$  0.19 \\
\hline
$\Delta \mu_{0}$\tablenotemark{b} & \nodata      &  \nodata          & 0.14 $\pm$ 0.02 & 0.04 $\pm$ 0.02 & 0.48  $\pm$ 0.02  & -0.07  $\pm$  0.10 & 0.01 $\pm$ 0.03 \\
$\chi^2$   &  \nodata     &  \nodata          &       2.7        &     1.0          &   3.5            &    2.6           &     1.6          \\
\enddata
\tablenotetext{a}{Weighted average of the I-band distance moduli from eq. \ref{eqbest1}-\ref{eqbest2}, and the V-band calibrations \ref{eqv}-\ref{eqggcv}.}
\tablenotetext{b}{$\Delta \mu_{0}=\mu_{0,cal}-\mu_{0,group}$, where $\mu_{0,cal}$ refers to the distance modulus obtained using
one of the calibrations \ref{eqt01}-\ref{eqm06}, together with  eq. \ref{eqggc} (see text).}
\label{tab_distances}
\end{deluxetable}

\clearpage

\bibliographystyle{apj}
\bibliography{cantiello_jun07}

\end{document}